\documentclass[aps,reprint,prappl,groupedaddress,superscriptaddress,showpacs,longbibliography]{revtex4-1}

\usepackage{natbib}
\usepackage[latin1]{inputenc}
\usepackage{graphicx}
\usepackage{amsmath,amssymb,stmaryrd,bm,dcolumn,textcomp}
\usepackage{mathrsfs}
\usepackage{rotating,array,color,float}
\usepackage{xcolor}
\usepackage{hyperref}					
\hypersetup{							    
    colorlinks,
    linkcolor={blue!80!black},
    citecolor={blue!80!black},
    urlcolor={blue!80!black}
}

\newcommand{\notop}{{{}_{}}}

\newcommand\Tstrut{\rule{0pt}{2.5ex}}         
\newcommand{\mr}[1]{\ensuremath{\mathrm{#1}}}
\renewcommand{\vec}[1]{\bm{#1}}
\newcommand{\ee}{\mathrm{e}}
\newcommand{\ii}{\mathrm{i}}
\newcommand{\dm}{\mathrm{d}}

\DeclareMathOperator{\re}{Re}
\DeclareMathOperator{\im}{Im}

\newcommand{\iot}{{\ii\omega t}}

\newcommand{\ve}{\varepsilon}

\newcommand{\pp}{\partial}       

\newcommand{\nablabf}{\boldsymbol{\nabla}}

\newcommand{\ddd}{\vec{d}}

\newcommand{\een}{\vec{e}^\notop}

\newcommand{\facp}{f^\notop_\mathrm{acp}}

\newcommand{\nhat}{\vec{\hat{n}}}

\newcommand{\qqq}{\vec{q}}

\newcommand{\rrr}{\vec{r}}

\newcommand{\uuu}{\vec{u}}

\newcommand{\zerovec}{\boldsymbol{0}}

\newcommand{\SIC}{\textrm{C}}

\newcommand{\SIkHz}{\textrm{kHz}}

\newcommand{\SIkgm}{\textrm{kg}\:\textrm{m$^{-3}$}}

\newcommand{\SIm}{\textrm{m}}

\newcommand{\SImum}{\textrm{\textmu{}m}}

\newcommand{\SIPa}{\textrm{Pa}}

\newcommand{\SIGPa}{\textrm{GPa}}

\newcommand{\SIs}{\textrm{s}}

\newcommand{\beq}[1]{\begin{equation} \eqlab{#1}}
\newcommand{\eeq}{\end{equation}}
\newcommand{\bsub}{\begin{subequations}}
\newcommand{\esub}{\end{subequations}}
\def\bal#1\eal{\begin{align}#1\end{align}}
\def\bsubal#1\esubal{\bsub \begin{align}#1\end{align} \esub}

\newcommand{\eqlab}[1]{\label{eq:#1}}
\renewcommand{\eqref}[1]{Eq.~(\ref{eq:#1})}

\newcommand{\eqnoref}[1]{(\ref{eq:#1})}

\newcommand{\figref}[1]{Fig.~\ref{fig:#1}}

\newcommand{\figlab}[1]{\label{fig:#1}}
\newcommand{\appref}[1]{Appendix~\ref{sec:#1}}

\newcommand{\secref}[1]{Section~\ref{sec:#1}}

\newcommand{\seclab}[1]{\label{sec:#1}}
\newcommand{\tabref}[1]{Table~\ref{tab:#1}}
\newcommand{\tabsref}[2]{Tables~\ref{tab:#1} and~\ref{tab:#2}}
\newcommand{\tablab}[1]{\label{tab:#1}}

\begin{document}

\title{Performance study of acoustophoretic microfluidic silicon-glass devices\\
by characterization of material- and geometry-dependent frequency spectra}

\author{Fabio Garofalo}
\email[corresponding author:]{fabio.garofalo@bme.lth.se}
\affiliation{Department of Biomedical Engineering, Lund University,
Ole R\"omers V\"ag 3 S-22363, Lund, Sweden}

\author{Thomas Laurell}
\email{thomas.laurell@bme.lth.se}
\affiliation{Department of Biomedical Engineering, Lund University,
Ole R\"omers V\"ag 3 S-22363, Lund, Sweden}

\author{Henrik Bruus}
\email{bruus@fysik.dtu.dk}
\affiliation{Department of Physics, Technical University of Denmark,\\ DTU Physics Building 309,
DK-2800 Kongens Lyngby, Denmark}


\begin{abstract}
The mechanical and electrical response of acoustophoretic microfluidic devices attached to an ac-voltage-driven piezoelectric transducer is studied by means of numerical simulations. The governing equations are formulated in a variational framework that, introducing Lagrangian and Hamiltonian densities, is used to derive the weak form for the finite element discretization of the equations and to characterize the device response in terms of frequency-dependent figures of merit or indicators. The effectiveness of the device in focusing microparticles is quantified by two mechanical indicators: the average direction of the pressure gradient and the amount of acoustic energy localized in the microchannel. Further, we derive the relations between the Lagrangian, the Hamiltonian and three electrical indicators: the resonance Q-value, the impedance and the electric power. The frequency response of the hard-to-measure mechanical indicators is correlated to that of the easy-to-measure electrical indicators, and by introducing optimality criteria, it is clarified to which extent the latter suffices to identify optimal driving frequencies as the geometric configuration and the material parameters vary. The latter have been varied by considering both pyrex and ALON top-lid.
\end{abstract}

%
\maketitle

\section{Introduction}
Based on the combined action of ultrasound waves and the flow of carrier fluids, acoustofluidics has emerged as a useful tool for manipulation of biofluids and biological suspensions in microfluidic devices.
These devices exploit standing acoustic pressure waves that through the purely mechanical parameters, such as compressibility, density and size, induce fluid- and particle-specific forces \cite{Bruus_2012_Acoustofluidics_7, Settnes_2012, Karlsen_2015} leading to acoustophoresis \cite{Bruus_2011}.
This phenomenon is the basis of the development of gentle \cite{Burguillos_2013, Wiklund_2012} and robust methods for concentrating \cite{Nordin_2012}, trapping \cite{Evander_2012}, washing \cite{Augustsson_2012}, aligning\cite{Manneberg_2008} and separating cells \cite{Augustsson_2012_Anal_Chem, Ding_2014, Petersson_2007}.
In order to be used for manipulation purposes, the acoustic pressure wave inside the microchannel must exhibit well-defined pressure nodes and intense pressure fields, that effectively attract or repel particles. For these reasons, acoustofluidic devices operate at acoustic resonance frequencies.
Because the speed of sound in water is around $1500~\mathrm{m/s}$, and the typical characteristic dimensions of acoustofluidic microchannels range in \mbox{$200-500~\SImum$}, it is seen that ultrasound frequencies of about \mbox{$1.5-2.5~\mathrm{MHz}$} are ideally suited for creating effective resonance conditions in acoustofluidic devices.

Despite the many successful acoustofluidic devices reported in the literature, a fair amount of calibration and fine tuning is still involved in the design, optimization and control during experiments, as to as properly identify the optimal working conditions.
While an experimental knowledge of the system response with respect to an external actuation source plays a central role in the selection of good operative conditions \cite{Hammarstrom_2014}, a better comprehension of the phenomena involved in these devices would free the design and implementation steps from the costly and time-consuming methods currently employed.

Recently, some progress have been made in numerical modeling of ultrasound and elastic waves in microscale acoustofluidic systems including the piezo transducer, chip as an elastic solid and the fluid inside the microchannels \cite{Friend_2011, Dual_and_Schwarz_2012, Dual_and_Moller_2012}.
Studies including the thermoviscous and transient effects inside microchannels have been reported as well \cite{Muller_2013, Muller_2014, Muller_2015}.
The first numerical optimization studies of acoustophoretic devices have also been performed recently, illustrating a procedure to obtain optimal acoustophoretic forces by changing the geometrical parameters of the device \cite{Hahn_2014}.
Other studies involve numerical characterization of the acoustic pressure wave in the microchannel and subsequent computation of particle trajectories by means of numerical integration \cite{Neild_2007, Oberti_2008, Muller_2012, Barnkob_2010, Augustsson_2011, Barnkob_2012}.

A major problem regarding the numerical optimization of acoustofluidic devices, is the lack of rigorous definitions of macroscopic descriptors, which (i)~characterize the efficiency of acoustophoretic devices for a given electrical actuation of the piezo transducer and (ii)~are accessible both experimentally and numerically. As an example, in \cite{Hahn_2014} an objective function, i.e. Eq.~(17), containing the Gorkov potential and therefore including both the device performance and the particle properties is used to optimize the performance in the space of the parameters. Note that since the objective function used in \cite{Hahn_2014} depends on the particle properties, it would be better to speak about ``separation performance for the X particle'' other than ``device performance''. In the present paper we are interested to investigate the device performance and thus we will introduce indicators that do not take into account for the particle properties but they regard exclusively the device characteristics.

Given the ease by which the electrical response of the piezo transducers can be measured compared to the acoustophoretic response, it would be advantageous to demonstrate numerically and experimentally a direct correlation between these two responses, such that the hard-to-obtain acoustophoretic response could be inferred from the easy-to-obtain electric response.
A first step in this direction has been taken in experiments on simple glass capillaries by Hammarstr\"om \textit{et al.} \cite{Hammarstrom_2014}.
A further complication is the influence of thermal heating by the piezo transducer, which directly affects the acoustic response of the device through the temperature dependence of the acoustic relevant parameters (density, compressibility and elastic moduli). However, because an overall temperature increase in an operative device can be easily prevented by including a Peltier cooling element attached to the device \cite{Augustsson_2011, Adams_2012}, we have chosen not to take thermal heating into account in the simulations here presented.

In order to characterize acoustophoretic devices, in this paper we aim 
(i)~to introduce the descriptors that enable for the quantitative analysis of frequency spectra in the case of purely electromechanical interactions of the device and the liquid-filled microchannel,
(ii)~to provide an easy-to-run two-dimensional model aiming to address the features of a fully three-dimensional device, and (iii)~to compare the behavior of different acoustofludic devices as a function of the geometrical and material parameters.

The manuscript is organized as follows. In \secref{model} (A)~we define the model device in the form of a microchannel embedded in a silicon substrate with a top lid and a piezo attached at the bottom of the silicon, (B)~we introduce the governing equations using the Lagrangian and Hamiltonian formalism, (C)~we formulate the coupling between the various sub-systems of the device, (D)~we introduce the system indicators, and (E)~we describe the finite-element discretization that constitutes the basis of the numerical simulations. The materials chosen for the top lid, either pyrex or ALON (Aluminium Nitroxide glass \cite{Surmet_2015}), are both transparent thus allowing for optical access to the microchannel.
In \secref{numerical} we report the details of the numerical implementation by (A)~illustrating the weak form used in the finite element discretization and (B)~validating the model in terms of global and frequency-wise errors, and energy consistency.
In \secref{results} (A)~we introduce the procedure to identify optimal acoustophoretic frequencies, (B)~we perform the mechanical spectral analysis of pyrex-silicon devices, (C)~we report some examples of identification of optimal frequencies, (D)~we perform electric spectral analysis of pyrex-silicon devices, and (E)~we introduce and validate a procedure to identify resonance frequencies from the impedance characteristics. In \secref{conclusion}, we discuss the results and address possible experimental tests. Finally, in \appref{2D3Danalysis} we discuss the meaning of the two-dimensional analysis and the extension of the actual model to a fully three-dimensional numerical model.

\section{Acoustophoretic Device Model}
\seclab{model}

\begin{figure}[!!t]
\includegraphics[width=\columnwidth]{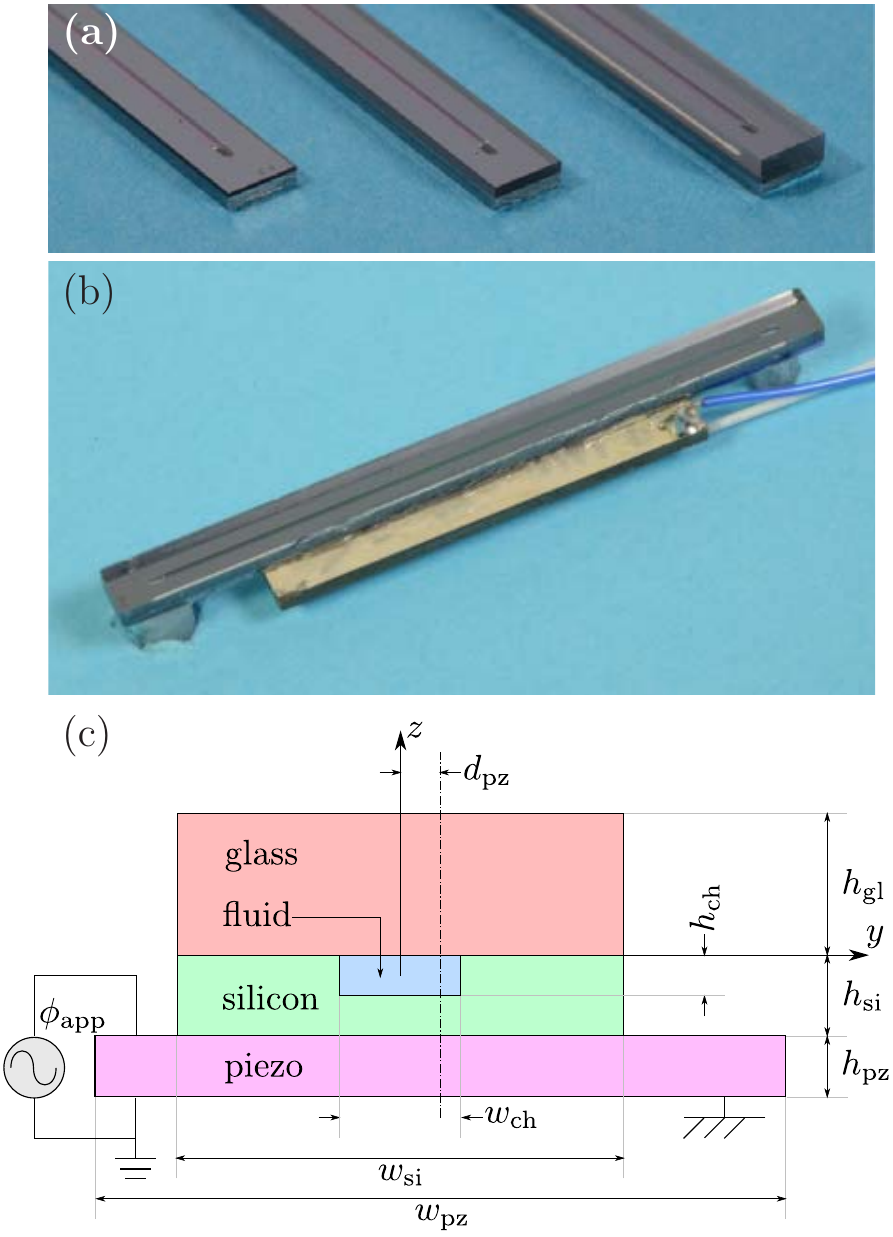}
\caption{\figlab{chipsANDdomain}
(a) Photograph of three prototypical acoustophoretic silicon-glass chips with glass lids of height 0.5, 0.7 and 1.1~mm, respectively. (b) Photograph of a complete device with a 40-mm long silicon-glass chip (gray) with short fluidic tubings attached underneath the chip ends and mounted on the piezo transducer (brown) connected by electrical wires (blue and white).
(c) The cross-section of the device defining the two-dimensional computational domain.
}
\end{figure}

\subsection{Description of the device}
The long and straight acoustophoretic device consists of a piezo transducer underneath a chip. The latter is made of a glass lid on top and a silicon substrate with an embedded microfluidic channel. In \figref{chipsANDdomain}, panels (a) and (b) show photographs of actual silicon-glass devices including the chip mounted on the piezo transducer, while panel (c) shows the cross-section of the device defining the computational domain used in the numerical study.
The piezoelectric transducer is modeled as a rectangular domain of width $w_\mathrm{pz}$ and height $h_\mathrm{pz}$ driven by an ac-voltage difference $\phi_\mr{app}$ applied between its top and bottom surfaces. The bottom surface is considered fixed, while the side surfaces move freely. The silicon substrate is a rectangular domain of width $w_\mathrm{si}$ and height $h_\mathrm{si}$. The rectangular microfluidic channel of width $w_\mathrm{ch}$ and height $h_\mathrm{ch}$ is etched into the top of the silicon substrate. Finally, a rectangular glass lid of width $w_\mathrm{gl} = w_\mathrm{si}$ and height $h_\mathrm{gl}$ is placed on top of the silicon substrate. The center line of the piezo transducer is displaced by a distance $d_\mathrm{pz}$ with respect to the center line of the chip.

\subsection{Governing equations and Lagrangian densities}
\seclab{GoverningEq}
In the following, we formulate the governing equations and report the free Lagrangian densities for the elastic, acoustic and piezoelectric waves in the case of time-harmonic actuation. The material properties are not specified, being tacitly assumed that these are different for different materials (see \secref{numerical} for further details).

\textit{Elastic waves.}
The propagation of elastic waves in the silicon chip and the glass lid is governed by the time-dependent Navier equation \cite{Landau_Elasticity, Graff_1991},
 \beq{tdneq}
 -\rho\partial_t^2\bm{u}+\nablabf \cdot\bm{\sigma}_\mathrm{m}
 = \zerovec\,, \end{equation}
where $\rho$ is the density of the material, $\bm{u}$ is the displacement and $\bm{\sigma}_\mathrm{m}$ is the mechanical stress tensor.
The constitutive equation relating the mechanical stress and displacement in the linear regime is given by
\cite{Landau_Elasticity, Chou_1992},
 \beq{leeq}
 \bm{\sigma}_\mathrm{m}=\bm{\Sigma}:\bm{\gamma}\,,
 \end{equation}
where $\bm{\gamma}=\frac{1}{2}[\nablabf \bm{u}+(\nablabf \bm{u})^{\mr{T}}]$ is the strain tensor, $\bm{\Sigma}$ is the rank-4 stiffness tensor. The symbols ``T" and $``\!:\!"$ denote the transpose and inner product of tensors, respectively. The stiffness tensor $\bm{\Sigma}$ has $3^4=81$ components, but the positivity of the elastic strain energy, expressed by the constraints $\Sigma_{ij,kl}=\Sigma_{kl,ij}$, $\Sigma_{ij,kl}=\Sigma_{ji,kl}$ and $\Sigma_{ij,kl}=\Sigma_{ij,lk}$, reduces the number of independent components to $21$. For mono-crystalline silicon, $\bm{\Sigma}$ is fully described by only three parameters, and for isotropic glass two parameters suffice (the Young modulus and the Poisson ratio) \cite{Graff_1991, Landau_Elasticity, Chou_1992}.

The Navier equation~\eqnoref{tdneq} can be written in the frequency-domain (Helmholtz form) by using the time-domain Fourier representation \mbox{$u(\bm{x},t)=u(\bm{x},\omega)e^{-i\omega\,t}$},
 \beq{neeq}
 \rho\,\omega^2\bm{u}+\nablabf \cdot\bm{\sigma}_\mathrm{m}=\bm{0}\,,
 \end{equation}
where $\omega=2\pi f$ is the angular frequency corresponding to the actuation frequency $f$ of the transducer.

In a variational approach to the field equations \cite{Moiseiwitsch_2004}, \eqref{neeq} is the Euler--Lagrange equation \mbox{$\pp \mathscr{L}_\mathrm{m}/\pp\uuu^* - \nablabf\cdot\big[\pp \mathscr{L}_\mathrm{m}/\pp(\nablabf\uuu^*)\big] = \zerovec$} derived from the variation of the mechanical Lagrangian \mbox{$L_m=\int \mathscr{L}_\mathrm{m}\,d\bm{x}$ with}
 \beq{mldeq}
 \mathscr{L}_\mathrm{m}(\bm{u},\nablabf \bm{u})=
 \rho\,\omega^2\bm{u}^*\cdot\bm{u}-
 \nablabf \bm{u}^*:\bm{\Sigma}:\nablabf \bm{u}\,,
 \end{equation}
where $``*"$ denotes the transposed conjugate.
Note that because $\bm{\Sigma}$ is symmetric in both
couples of its indices, the inner product
\mbox{$\nablabf \bm{u}^*:\bm{\Sigma}:\nablabf \bm{u}$} equals
\mbox{$\bm{\gamma}^*:\bm{\Sigma}:\bm{\gamma}$}.
This rule will be
used in the following to disregard the anti-symmetric part of the deformation
$\nablabf \bm{u}$ when it is double-dotted with a symmetric tensor, such as $\bm{\Sigma}$.

\textit{Acoustic waves.}
The governing equation for the acoustic waves in the frequency-domain (inviscid fluid) inside the microfluidic channel is the wave equation \cite{Morse_1987},
 \beq{tdaweq}
 \kappa_s p + \frac{1}{\rho\omega^2} \nablabf \cdot \nablabf p=0\,,
 \end{equation}

where $\rho$ is the density of the fluid, \mbox{$\kappa_s=\frac{1}{\rho\,c^2}$} is its isentropic compressibility, $c$ the speed of sound (considered complex-valued as to include bulk dissipation), and $p$ is the acoustic pressure.
The governing equation can be derived from the acoustic Lagrangian density $\mathscr{L}_\mathrm{a}$ of the fluid,
 \beq{aldeq}
 \mathscr{L}_\mathrm{a}(p,\nablabf p)= \kappa_s\,p^*p
 - \frac{1}{\rho\,\omega^2} \nablabf p^*\cdot\nablabf p\,.
 \end{equation}
Note that (i)~the factor $\omega^2$ in \eqref{tdaweq} appears in such a way that the Lagrangian $\mathscr{L}_\mathrm{a}$ has the dimension of an energy density, the same as $\mathscr{L}_\mathrm{m}$, and (ii)~the role of the kinetic and potential energies in $\mathscr{L}_\mathrm{a}$ is reversed when compared to $\mathscr{L}_\mathrm{m}$, since in acoustics the potential energy is the $p^*p$-term and the kinetic energy is the gradient-product-term \cite{Goldstein_2001, Morse_1987, Xu_2011}.

\textit{Piezoelectric Waves.}
The governing equations in the linear deformation regime for the electromechanical (em) or piezoelectric waves in the piezo transducer are the Helmholtz--Navier equation~\eqnoref{neeq} coupled with the conservation for the electric displacement $\ddd_\mathrm{em}$ in absence of electric charge distribution\cite{Qin_2012},
 \bsub
 \eqlab{tdpeq}
 \bal
 \eqlab{tdpeq_a}
 \rho\,\omega^2\bm{u}+\nablabf \cdot\bm{\sigma}_\mathrm{em}&=\bm{0}\,,\\
 \eqlab{tdpeq_b}
 \nablabf \cdot\bm{d}_\mathrm{em}&=0\,,
 \eal
 \esub
where $\bm{\sigma}_\mathrm{em}$ is the electromechanical stress tensor, and
$\bm{d}_\mathrm{em}$ is the electric displacement vector.
The linear stress-charge constitutive equations \cite{Qin_2012} for $\bm{\sigma}_\mathrm{em}$ and $\ddd_\mathrm{em}$ are extensions of \eqref{leeq} and the standard constitutive relation for the electric displacement,
 \bsub
 \eqlab{peceq}
 \begin{align}
 \eqlab{peceq_a}
 \bm{\sigma}_\mathrm{em}&=\bm{\Sigma}:\bm{\gamma}+\bm{P}^\dag\cdot\nablabf \phi\,,\\
 \eqlab{peceq_b}
 \bm{d}_\mathrm{em}&=\bm{P}:\bm{\gamma}-\bm{\varepsilon}\cdot\nablabf \phi\,,
 \end{align}
 \esub
where $\phi$ is the electrostatic potential, and $\bm{\varepsilon}$ is the dielectric permittivity tensor.
The piezoelectric coupling tensor $\bm{P}$ is a third-order tensor, that is symmetric in its rightmost indices, \mbox{$P_{i,jk}=P_{i,kj}$}, and where \mbox{$[\bm{P}^\dag]_{kj,i} = [\bm{P}]^*_{i,jk}$} is the Hermitian conjugate.
These symmetries imply that \mbox{$\bm{P}:\bm{\gamma} = \bm{P}:\nablabf \bm{u}$}.
Finally, the corresponding Lagrangian density $\mathscr{L}_\mathrm{em}$ is
 \begin{align}
 \eqlab{emleq}
 \mathscr{L}_\mathrm{em}(\bm{u},\nablabf \bm{u},\phi,\nablabf \phi)
 &=\rho\,\omega^2\bm{u}^*\cdot\bm{u}
 -\nablabf \bm{u}^*:\bm{\Sigma}:\nablabf \bm{u}
 \nonumber\\
 &\quad
 -\nablabf \phi^*\!\cdot\!\bm{P}:\nablabf \bm{u}
 -\nablabf \bm{u}^*\!:\!\bm{P}^\dag\!\cdot\nablabf \phi
 \nonumber \\
 &\quad+\nablabf \phi^*\cdot\bm{\varepsilon}\cdot\nablabf\phi\,,
 \end{align}
that is similar to that provided by \cite{Qin_2012} (discussion and table in Sec.~1.2) for the static case, except for the fact that the kinetic term, i.e. \mbox{$\rho\,\omega^2\bm{u}^*\cdot\bm{u}$}, has been included.

\subsection{Boundary conditions}
\seclab{BoundaryCond}

The boundary conditions for the above-mentioned four domains (piezo transducer, silicon chip, water channel and glass lid) are set as follows.

Mechanically, the piezo transducer is free to vibrate (zero stress) at the bottom wall $S_\mathrm{pz/bot}$ and at the interface $S_\mathrm{pz/air}$, while the displacement and the stress are continuous across the interface $S_\mathrm{pz/si}$ with the silicon chip,
 \bsubal
 \eqlab{BC_pzM_bot}
 \bm{\sigma}_\mathrm{em}\cdot\nhat&=\bm{0}\,,\quad \bm{x}\in S_\mathrm{pz/bot}\,,\\
 \eqlab{BC_pzM_air}
 \bm{\sigma}_\mathrm{em}\cdot\nhat&=\bm{0}\,,\quad \bm{x}\in S_\mathrm{pz/air}\,,\\
 \eqlab{BC_pzM_si}
 \llbracket\bm{u}\rrbracket=\bm{0} \text{ and }
 \llbracket\bm{\sigma}\rrbracket\cdot\nhat&=\bm{0},\quad\bm{x}\in S_\mathrm{pz/si}\,,
 \esubal
where \mbox{$\llbracket g\rrbracket=g_2-g_1$} is the difference in $g$ across an interface separating ``2" from ``1" (note that the stress in the piezo is $\bm{\sigma}_\mathrm{em}$ as defined in \eqref{emleq}, while in the silicon is $\bm{\sigma}_\mathrm{m}$ as defined in \eqref{leeq}). Electrically, the piezo transducer has zero potential at the bottom wall $S_\mathrm{pz/bot}$, an externally applied potential $\phi_\mathrm{app}$ on the top wall $S_\mathrm{pz/top}$, and zero surface charge on the side walls $S_\mathrm{pz/side}$,
 \bsubal
 \eqlab{BC_pzE_bot}
 \phi&=0\,,\quad \bm{x}\in S_\mathrm{pz/bot}\,,\\
 \eqlab{BC_pzE_top}
 \phi-\phi_\mathrm{app}&=0\,,\quad \bm{x}\in S_\mathrm{pz/top}\,,\\
 \eqlab{BC_pzE_side}
 \bm{d}_\mathrm{em}\cdot\nhat &= 0\,,\quad\bm{x}\in S_\mathrm{pz/side}\,.
 \esubal
Given the linearity of the model here presented, we can assume, without loss of generality, that the applied voltage is \mbox{$\phi_\mathrm{app} = 1\,\mathrm{V}$} (for real-world devices this voltage is usually between 3-7 Volts).

The remaining boundary conditions are purely mechanical. The glass and silicon surfaces exposed to air ($S_\mathrm{gl/air}$ and $S_\mathrm{si/air}$) have zero stress, the silicon-glass interface ($S_\mathrm{si/gl}$) has continuous displacement field and stress tensor, while the glass and silicon surfaces ($S_\mathrm{gl/fl}$ and $S_\mathrm{si/fl}$) exposed to the fluid in the microchannel have continuous stress (the tangential stress in the inviscid fluid is by definition zero) and normal displacement,
 \bsubal
 \eqlab{BC_siM_air}
 \bm{\sigma}\cdot\nhat&=\bm{0}\,,\quad\bm{x}\in S_\mathrm{si/air}, S_\mathrm{gl/air},\\
 \eqlab{BC_siM_gl}
 \llbracket\bm{u}\rrbracket=\bm{0} \text{ and }
 \llbracket\bm{\sigma}\rrbracket\cdot\nhat&=\bm{0}\,,\quad\bm{x}\in S_\mathrm{si/gl}\,,\\
 \eqlab{BC_siM_fl}
 \llbracket\bm{u}\rrbracket\cdot\nhat=0 \text{ and }
 \llbracket\bm{\sigma}\rrbracket\cdot\nhat&=\bm{0}\,,
 \quad\bm{x}\in S_\mathrm{si/fl}\,, S_\mathrm{gl/fl}.
 \esubal
Note that for the fluid, the displacement field is given by \mbox{$\bm{u}_\mathrm{f} = \frac{1}{\rho\omega^2}\:\nablabf p$} and the stress by \mbox{$\bm{\sigma}_\mathrm{f} = - p\bm{1}$}, where $\bm{1}$ is the unit tensor.

So far, the boundary conditions have been expressed in their strong forms. However for the numerical implementation in the finite-element method, we need to formulate them in weak form, and adding the contributions so obtained to the free Lagrangian densities Eqs.~\eqnoref{mldeq}, \eqnoref{aldeq} and \eqnoref{emleq} leads to correctly constrained solutions to the variational problem.

The boundary terms for the free Lagrangian densities $\mathscr{L}_\mathrm{m}$, $\mathscr{L}_\mathrm{a}$ and $\mathscr{L}_\mathrm{em}$, are derived from the virtual work form of Eqs.~\eqnoref{neeq}, \eqnoref{tdaweq} and \eqnoref{peceq} by considering integration over the relevant domain and application of the Green-Gauss theorem.
This operation yields the sought contributions in terms of the surface-normal component of the fluxes,
 \bsub
 \eqlab{lagr_dens_bnd}
 \begin{align}
 \mathscr{L}_\mathrm{m}^\mathrm{bnd}(\bm{u},\bm{\sigma}_\mathrm{m}^\mathrm{bnd}
 )&=
 \bm{u}^*\cdot\bm{\sigma}_\mathrm{m}^\mathrm{bnd}\cdot\bm{\hat{n}}\,,\\
 \mathscr{L}_\mathrm{a}^\mathrm{bnd}(p,\bm{u}^\mathrm{bnd}_\mathrm{f})&=
 p^*\,\bm{u}^\mathrm{bnd}_\mathrm{f}\cdot\bm{\hat{n}}\,,\\
 \mathscr{L}_\mathrm{em}^\mathrm{bnd}
 (\bm{u},\bm{\sigma}_\mathrm{em}^\mathrm{bnd},\phi,
 \bm{d}_\mathrm{em}^\mathrm{bnd})&=\bm{u}^*\cdot\bm{\sigma}_\mathrm{em}
 ^\mathrm{bnd}
 \cdot\bm{\hat{n}}-
 \phi^*\,\bm{d}_\mathrm{em}^\mathrm{bnd}\cdot\bm{\hat{n}},
 \end{align}
 \esub
where the reaction forces, the dot product of the fluxes with the surface normal $\nhat$, are explicitly addressed. At interfaces between neighboring domains, these contributions cancel due to the presence of surface normals that have opposite direction in the two domains. Therefore, if no extra surface contribution is added to the Lagrange densities, the continuity condition $\llbracket\cdot\rrbracket = 0$ for the normal fluxes at such interfaces is automatically fulfilled.
For free boundaries, such as external walls, the reaction forces are by definition zero, and if no further terms are added to the Lagrangian, these boundary conditions are also fulfilled.

On the bottom wall of the piezo transducer, the Dirichlet condition on the electrical potential $\phi$ are imposed using the standard method of Lagrange multiplier functions, here $\mu_\mathrm{em}$, respectively. This corresponds to add the following Lagrange surface density contribution,
 \begin{equation}
 \mathscr{L}_\mathrm{em}^\mathrm{bot}=
 \phi^*\mu_\mathrm{em}+
 \mu_\mathrm{em}^*(-\phi)\,,
 \end{equation}
to the free Lagrange mechanical density \mbox{$\mathscr{L}_\mathrm{em}^{}$}, through which we obtain the Euler--Lagrange equation $\pp_{\mu_\mathrm{em}^*} \mathscr{L}_\mathrm{em}^\mathrm{bot} = 0$, which implies $\phi = 0$. Similarly, for the piezo top wall we add the following Lagrange surface density $\mathscr{L}_\mathrm{em}^\mathrm{top}$ to the free Lagrange mechanical density,
 \beq{consttop}
 \mathscr{L}_\mathrm{em}^\mathrm{top}=
 \phi^*\mu_\mathrm{em}+\mu_\mathrm{em}^*(-\phi+\phi_\mathrm{app}).
 \end{equation}
the Euler--Lagrange equation for $\mu_\mathrm{em}^*$ leads to the correct boundary condition $\phi = \phi_\mathrm{app}\:\ee^{-\iot}$ on the top wall.

By using the virtual work theorem and the Gauss's theorem, it is straightforward to prove that the total work \mbox{$\hat{W}(\omega)=\int_{S_\mathrm{top}}\mathscr{W}(\bm{x},\omega)\:\dm S$} made by the external actuation source on the acoustofluidic device equals the volume integral over the system domain $\Omega$ of the total Lagrange density
\mbox{$\mathscr{L}=\mathscr{L}_\mathrm{m}+\mathscr{L}_\mathrm{a}+\mathscr{L}_\mathrm{em}$},
 \beq{lwequiv}
 \hat{W}(\omega)= \int_\Omega\mathscr{L}(\bm{x},\omega)\: \dm\bm{x}.
 \end{equation}
Henceforth, we use the hat-notation on capital letters to denote complex-valued
integral quantities such as $\hat{W}(\omega)$.

\subsection{Definition of system indicators}
\seclab{encharsec}
From the Lagrangian densities introduced in the previous section, we can derive the expressions for the Hamiltonian densities, the stored energy and the dissipated power for the device. By using complex notation and treating a field $\bm{q}$ and its complex conjugate $\bm{q}^*$ as independent variables, the corresponding Hamiltonian density is given by \mbox{$\mathscr{H} = \bm{q}^*\cdot\frac{\pp \mathscr{L}}{\pp \bm{q}^*} + \bm{q}\cdot\frac{\pp \mathscr{L}}{\pp \bm{q}} - \mathscr{L}$}. This prescription results in the following three Hamiltonian densities for each of the three subsystems,
 \bsub
 \eqlab{hdeq}
 \begin{align}
 \eqlab{hdeq_a}
 \mathscr{H}_\mathrm{m}&=\rho\,\omega^2\bm{u}^*\cdot\bm{u} +
 \nablabf \bm{u}^*:\bm{\Sigma}:\nablabf \bm{u},
 \\
 \eqlab{hdeq_b}
 \mathscr{H}_\mathrm{a} &= \kappa_s\,p^*p
 + \frac{1}{\rho\,\omega^2}\nablabf p^*\cdot\nablabf p,
 \\
 \eqlab{hdeq_c}
 \mathscr{H}_\mathrm{em} &=
 \rho\,\omega^2\bm{u}^*\cdot\bm{u}
 +\nablabf \bm{u}^*:\bm{\Sigma}:\nablabf \bm{u}
 - \nablabf \phi^*\cdot\bm{\varepsilon}\cdot\nablabf\phi
 \nonumber \\
 &\quad +\nablabf \phi^*\!\cdot\!\bm{P}:\nablabf \bm{u}
 -\nablabf \bm{u}^*\!:\!\bm{P}^\dag\!\cdot\nablabf \phi.
 \end{align}
 \esub
Note that to interpret the quantities introduced in \eqref{hdeq} as energies, the real part of these quantities must be positive. This (i) implies constraints on the actual material parameters (in terms of numerical values and symmetries of the higher-order tensor) and (ii) provides a further method to check the energy consistency of the model here presented.

The quadratic structure of the Hamiltonian densities implies that for a given domain $n =$ pz, si, gl, or ch, the contribution $H_n(\omega)$ to the total energy of the system is given by the time average over one oscillation period of the integrated complex-valued Hamiltonian $\hat{H}_n(\omega)$ as
 \bsub
 \bal
 \eqlab{Hn_contrib}
 H_n(\omega) &= \frac{1}{2}\re\big[\hat{H}_n(\omega)\big], \\
 \hat{H}_n(\omega) &= \int_{\Omega_n} \mathscr{H}_n(\bm{x},\omega)\:\dm \bm{x},
 \quad n = \text{ pz, si, gl, ch}.
 \eal
 \esub
The total energy per cycle $H(\omega)$ of the system is thus
 \beq{Htotal}
 H(\omega) = H_\mr{pz}(\omega) + H_\mr{si}(\omega)
 + H_\mr{si}(\omega) + H_\mr{ch}(\omega).
 \eeq

As described in \eqref{consttop} the system is driven by the applied potential \mbox{$\phi = \phi_\mathrm{app}\:\ee^{-\iot}$} on the top wall of the piezo. The dissipated power $P(\omega)$ is thus given by the time average over one period of the complex-valued rate \mbox{$\hat{P}=-\ii \omega  \hat{W}(\omega)$} of the applied work $\hat{W}(\omega)$ given in \eqref{lwequiv},
 \beq{Power}
 P(\omega) = \frac{1}{2} \re\big[\hat{W}(\omega)\big]
 = \frac{\omega}{2} \im\big[\hat{L}(\omega)\big],
 \eeq
where the total complex-valued Lagrangian $\hat{L}(\omega)$ of the system in analogy with the Hamiltonian is given by
 \bsub
 \bal
 \eqlab{Ltotal}
 \hat{L}(\omega) &= \hat{L}_\mr{pz}(\omega) + \hat{L}_\mr{si}(\omega)
 + \hat{L}_\mr{si}(\omega) + \hat{L}_\mr{ch}(\omega),
 \\
 \eqlab{Ln_contrib}
 \hat{L}_n(\omega) &= \int_{\Omega_n} \mathscr{L}_n(\bm{x},\omega)\:\dm \bm{x},
 \quad n = \text{ pz, si, gl, ch}.
 \eal
 \esub

An important indicator for characterizing the system in terms of electric measurements, is the complex-valued electrical impedance $\hat{Z}(\omega)$ defined through the voltage-current
relation $\phi_\mathrm{app} = \hat{Z}\hat{I}$. Multiplying this expression by $\phi_\mathrm{app}^*$, and recalling that the dissipated power is given by $\hat{P}=\phi_\mathrm{app}^*\hat{I}=-i\omega\hat{L}$, we obtain the electrical impedance,
 \beq{impdef}
 \hat{Z}(\omega)=i\frac{|\phi_\mathrm{app}|^2}{\omega \hat{L}(\omega)}\,,
\qquad \hat{Z}=Z(\omega)\exp[i\,\varphi_Z(\omega)]\,,
 \eeq
where $Z$ is the absolute value of the impedance and $\varphi_Z$ is the phase.
Since $\hat{L}(\omega)$ depends quadratically on the applied potential, we note that as expected the electrical impedance is independent of $\phi_\mathrm{app}$. Below we investigate to which extent the electrical impedance, which is easy to measure, characterizes the acoustic part of the electro-mechanical resonances of the system.

Additionally, from the total Hamiltonian $H(\omega)$ and the dissipated power $P(\omega)$, one can calculate the quality factor or Q-value for the electrically driven acoustofluidic device as \cite{Nilsson_2014},
 \beq{qfdef}
 Q(\omega) = \frac{\omega H(\omega)}{P(\omega)}
 = \frac{\re\big[\hat{H}(\omega)\big]}{\im\big[\hat{L}(\omega)\big]}.
 \eeq
The Q-value is one of the indicators used to characterize acoustofluidic chips,
indeed high Q-values indicate strong and well-defined resonances of the system.
However, we drop this in favour of an electric characterization based on the impedance \eqref{impdef}.

The above indicators characterize the system in its entirety.
An indicator addressing indirectly the amount of radiation force the system is capable to produce within the microchannel, is the fraction $\eta$ of the total energy $H(\omega)$ that resides as acoustic energy $H_\mathrm{ch}(\omega)$ in the channel,
 \beq{etaDef}
 \eta(\omega)=\frac{H_\mathrm{ch}(\omega)}{H(\omega)}\,.
 \eeq
Indeed, the acoustic radiation force responsible for the acoustophoresis in the channel is proportional to $H_\mathrm{ch}$ \cite{Bruus_2012_Acoustofluidics_7} and thus to $\eta$.
As $\eta$ approaches unity, the acoustic radiation force in the fluid attains the maximum radiation force achievable for a device as all the energy is stored as acoustic energy in the fluid. According to this property, $\eta$ can be termed as the \textit{acoustofluidic yield}.

However, the acoustofluidic yield only quantifies the magnitude of the acoustophoretic forces. To obtain good acoustophoresis, say particle separation, also the direction of these forces needs to be optimal. To quantify this, we introduce the \textit{acoustophoretic mean orientation} $\theta(\omega)$, defined in terms of the sine of the average direction of the pressure gradient (horizontal/vertical),
 \beq{mopw}
 \theta(\omega) = \sin\Bigg[
 \arctan\frac{\left\|\partial_y p(\bm{x},\omega)\right\|}
 {\left\|\partial_z p(\bm{x},\omega)\right\|}\Bigg].
 \eeq
Here, the average $\left\| f(\bm{x}) \right\|$ of a field $f(\bm{x})$ is defined as the average of the absolute value $|f(\bm{x})|$ in the channel domain $\Omega_\mr{ch}$.

Good acoustophoresis, namely focusing by horizontal motion towards the vertical center plane, is obtained when the average of the horizontal pressure gradient $\|\pp_y p\|$ is much larger than the vertical one $\|\pp_z p\|$. For example, in the ideal case of a horizontal standing pressure half-wave, \mbox{$p(\bm{x}) = p_0 \cos\big[\pi\frac{y}{w_\mathrm{ch}}\big]$}, we obtain \mbox{$\pp_z p = 0$} and thus \mbox{$\theta = 1$}. In contrast, a dominating vertical gradient corresponding to \mbox{$\theta = 0$} is useless. The acoustophoretic mean orientation $\theta$ contains only information about the direction of the pressure wave.
If information  is required about other features of the pressure field (number and distribution of the nodes and/or spatial homogeneity), additional indicators must be introduced. For the analysis presented below, the use of $\theta$ suffices.

It must be noted that the acoustophoretic mean orientation is based
on the assumption that the dipole scattering coefficient in the Gorkov
potential\cite{Settnes_2012} equals zero, since the definition
\eqref{mopw} takes into account just for the gradient of the pressure
field. This is a good approximation when the acoustophoretic separation
involves cells suspended in aqueous media, as the cell density is usually
quite close to water density, almost canceling the contribution of
the dipole scattering coefficient.

\section{Numerical implementation}
\seclab{numerical}
\subsection{Finite element discretization}
The Lagrangian representation introduced in \secref{GoverningEq} and the corresponding boundary contributions in \secref{BoundaryCond}, are suitable for implementing numerical simulations with finite element method.
The numerical simulations are performed using the finite element software COMSOL \cite{Comsol44_2013}. The coarse mesh shown in \figref{mesh} is generated by COMSOL. A finer mesh is used in the actual simulations and generated by assuming the geometric parameters in \tabref{geometry}.

\begin{figure}[!!b]
\includegraphics[width=\columnwidth]{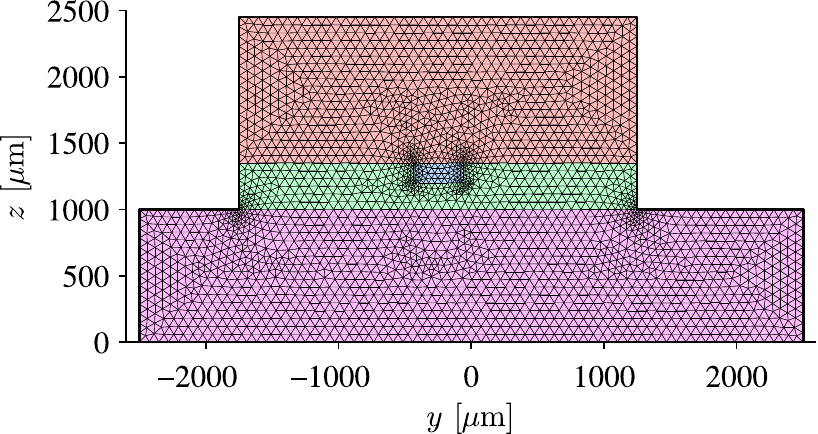}
\caption{\figlab{mesh}
Example of a mesh generated in the computational domain defined in \figref{chipsANDdomain} with a mesh resolution $m_\mr{res} = 0.35$, see details in \secref{validation}. The geometry parameters are listed in \tabref{geometry}. Specifically, the glass lid height is $h_\mathrm{gl}=1100~\SImum$ and the displacement is $d_\mathrm{pz}=250~\SImum$.}
\end{figure}

\begin{figure}[!!t]
\includegraphics[width=\columnwidth]{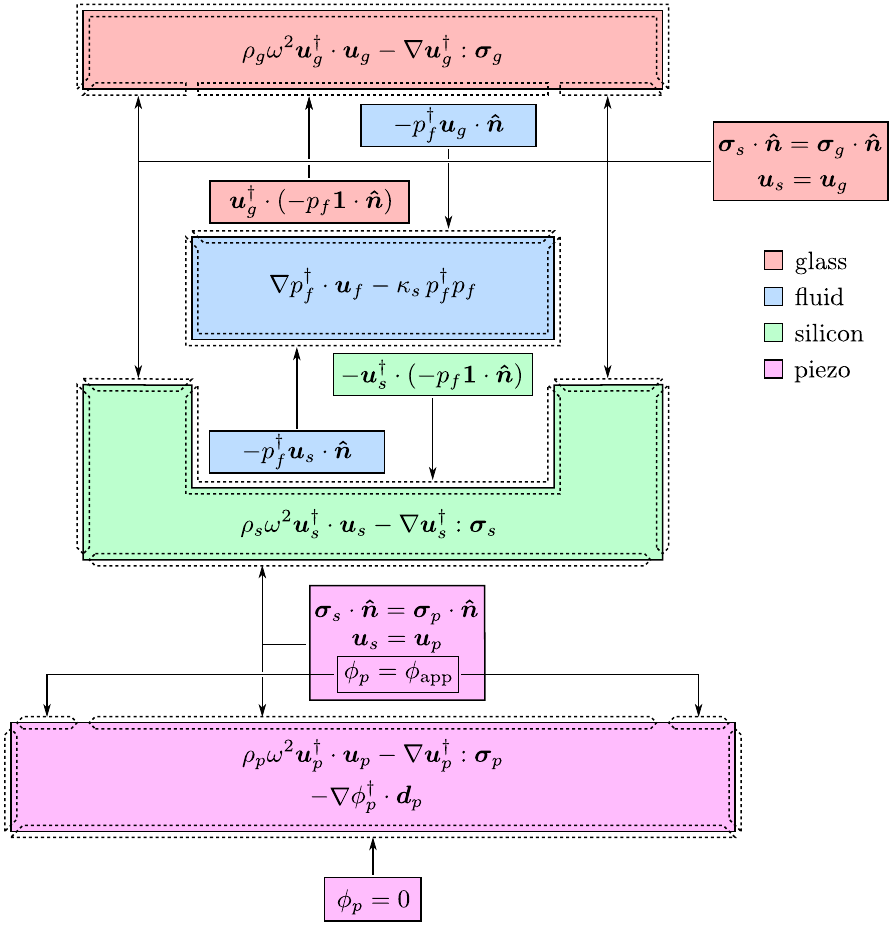}
\caption{\figlab{fem_scheme}
Block diagram illustrating the weak-form implementation of the bulk equations of \secref{GoverningEq} (larger boxes) and the boundary conditions of \secref{BoundaryCond} (smaller boxes). The system is driven by the applied potential $\phi_\mr{app}$ at angular frequency $\omega$. The fields are displacement $\bm{u}$, stress $\bm{\sigma}$, pressure $p$, electric potential $\phi$ and electric displacement $\bm{d}$. The material parameters are listed in \tabref{parameters}. The subscripts refer to the sub-domains.}
\end{figure}

\begin{table}[!!t]
\caption{\tablab{geometry} List of the geometry parameters for glass, silicon, channel and Pz26 used in the simulations.}
\begin{ruledtabular}
\begin{tabular}{lcl}
Length & Symbol & Value \Tstrut \\
\hline
 Glass height   & $h_\mathrm{gl}$ & $200, 500, 700, 1100\:\SImum$ \Tstrut\\
 Glass width    & $w_\mathrm{gl}$ & $375\:\SImum$\\
 Silicon height & $h_\mathrm{si}$ & $350\:\SImum$\\
 Silicon width  & $w_\mathrm{si}$ & $3000\:\SImum$\\
 Channel height & $h_\mathrm{ch}$ & $150\:\SImum$\\
 Channel width  & $w_\mathrm{ch}$ & $375\:\SImum$\\
 Pz26 height & $h_\mathrm{pz}$ & $1000\:\SImum$\\
 Pz26 width  & $w_\mathrm{pz}$  & $5000\:\SImum$\\
 Pz26 displacement   & $d_\mathrm{pz}$ & $0, 250, 500\:\SImum$\\
\end{tabular}
\end{ruledtabular}
\end{table}	

The Lagrangian densities~\eqnoref{mldeq}, \eqnoref{aldeq} and \eqnoref{emleq} corresponding to the governing equations \eqnoref{neeq}, \eqnoref{tdaweq} and \eqnoref{tdpeq}, are implemented in COMSOL by substituting the complex conjugate fields $\bm{u}^*$, $p^*$ and $\phi^*$ by the so-called test functions $\bm{u}^\dagger$, $p^\dagger$ and $\phi^\dagger$. For the sake of clarity, an overview of the weak-form implementation in COMSOL is shown in the block diagram of \figref{fem_scheme}. The larger boxes in the diagram contain the sub-domain bulk equations of \secref{GoverningEq}, while the boundary conditions of \secref{BoundaryCond} are displayed in the smaller boxes from which arrows point to the interfaces (dashed regions), where they have been applied. Conditions Eq.~\eqnoref{BC_siM_fl} are implemented by the substitutions
$\bm{\sigma}_\mr{s} \rightarrow -p_\mr{f} \bm{1}$ in $\mathscr{L}_\mr{m}^\mathrm{bnd}$ and
$\bm{u}_\mr{f} \rightarrow \bm{u}_\mr{s}$ in $\mathscr{L}_\mr{f}^\mathrm{bnd}$. \tabsref{geometry}{parameters} contain geometrical and material parameters used in the simulations.

\begin{table}[t]
\caption{\tablab{parameters} List of the material parameters used in the simulations. We assume $Q_\mathrm{si} = \infty$ and $Q_\mathrm{al} = Q_\mathrm{py}$.}
\begin{ruledtabular}
\begin{tabular}{ll}
Parameter & Symbol and value \Tstrut
\\ \hline
\multicolumn{2}{l}{\textit{Water parameters:} \cite{Muller_2014, Hahn_2014}} \Tstrut \\
 Speed of sound & $c_\mathrm{wa} = \big(1+i\frac12 \varphi_\mathrm{wa}\big)\:1481\:\SIm\,\SIs^{-1}$\\
 Density & $\rho_\mathrm{wa}=998\:\SIkgm$\\
 Loss factor & $\varphi_\mathrm{wa} = 0.01$\\[1.5mm]
\multicolumn{2}{l}{\textit{Pyrex parameters:} \cite{Schott_2015}}\\
 Density & $\rho_\mathrm{py} =2220\:\SIkgm$\\
 Young Modulus & $E_\mathrm{py} = \big(1+i\frac{1}{Q_\mathrm{py}}\big)\:63\:\SIGPa$\\
 Poisson ratio & $\nu_\mathrm{py} = 0.2$\\
 Quality factor & $Q_\mathrm{py} = 1250$\\[1.5mm]
\multicolumn{2}{l}{\textit{ALON parameters:} \cite{Surmet_2015}}\\
 Density & $\rho_\mathrm{al} = 3688\:\SIkgm$\\
 Young Modulus & $E_\mathrm{al} = \big(1+i\frac{1}{Q_\mathrm{al}}\big)\:334\:\SIGPa$\\
 Poisson ratio & $\nu_\mathrm{al} =0.239$\\
 Quality factor & $Q_\mathrm{al} = 1250$\\[1.5mm]
\multicolumn{2}{l}{\textit{Silicon parameters:} \cite{Dual_and_Moller_2012, Hopcroft_2010}}\\
Density & $\rho_\mathrm{si} =2330\:\SIkgm$\\
Stiffness matrix & $\bm{\Sigma}_\mathrm{si} =\bm{\Sigma}'$, \text{with elements}\\
 & $\Sigma'_{11}=\Sigma'_{22}=\Sigma'_{33}=165.7\,\SIGPa$\\
 & $\Sigma'_{12}=\Sigma'_{13}=63.9\:\SIGPa$\\
 & $\Sigma'_{44}=\Sigma'_{55}=\Sigma'_{66}=79.6\:\SIGPa$\\[1.5mm]
\multicolumn{2}{l}{\textit{Pz26 parameters:} \cite{Ferroperm_2015}}\\
 Density & $\rho_\mathrm{pz} =7700\:\SIkgm$\\
 Stiffness matrix & $\bm{\Sigma}_\mathrm{pz} =
 \big(1+i\frac{1}{Q_\Sigma}\big)\:\bm{\Sigma}'$, \text{with elements}\\
 & $\Sigma'_{11}=\Sigma'_{22}=168\:\SIGPa$\\
 & $\Sigma'_{12}=110\:\SIGPa,\; \Sigma'_{13}=99.9\:\SIGPa$\\
 & $\Sigma'_{23}=99.9\:\SIGPa, \Sigma'_{33}=123\:\SIGPa$\\
 & $\Sigma'_{44}=\Sigma'_{55}=30.1\:\SIGPa$\\
 & $\Sigma'_{66}=28.8\:\SIGPa$\\
 Quality factor for $\Sigma$ & $Q_\Sigma = 100$\\
 Dielectric tensor & $\bm{\varepsilon}= \big(1-\frac{1}{Q_\ve}\big)\:\bm{\varepsilon}$\\
 & $\varepsilon_{11}=\varepsilon_{22}=828$\\
 & $\varepsilon_{33}=700$\\
Quality factor for $\ve$ & $Q_\ve = 333$\\
Coupling Matrix &$\bm{P}_\mathrm{pz} =
  \big(1+i\frac{1}{Q_\mathrm{pz}}\big)\:\bm{P}'$, \text{with elements}\\
 & $P'_{15}=P'_{24}=9.86\:\SIC\,\SIm^{-2}$\\
 & $P'_{31}=P'_{32}=-2.8\:\SIC\,\SIm^{-2}$\\
 & $P'_{33}=14.7\:\SIC\,\SIm^{-2}$\\
 Quality factor  & $Q_\mathrm{pz} = \frac{2Q_\Sigma Q_\ve}{Q_\ve-Q_\Sigma} = 286$\\
\end{tabular}
\end{ruledtabular}
\end{table}	

\begin{figure}[!!t]
\hspace*{-1.2cm}\centerline{\includegraphics[]{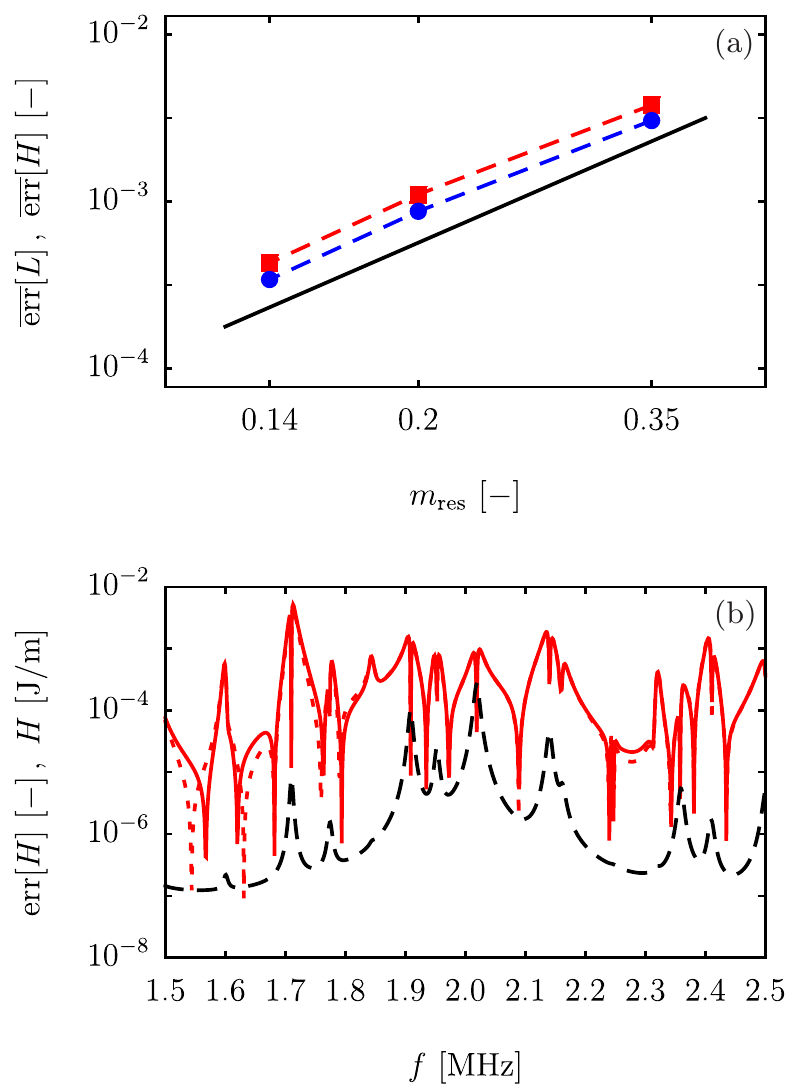}}
\caption{\figlab{errorfig}(a) Log-log plot of
$\overline{\mathrm{err}}\big[L(m_\mathrm{res})\big]$ ($\bullet$) and
$\overline{\mathrm{err}}\big[H(m_\mathrm{res})\big]$ ({\tiny $\blacksquare$}). The straight line is a reference slope corresponding to $m_\mathrm{res}^2$.
(b) Plot of the relative error for the real part (red full line), the
imaginary part (red dashed line) of the Hamiltonian (black dashed line) for $m_\mr{res} = 0.14$.}
\end{figure}

\subsection{Model validation}
\seclab{validation}
We have performed a number of tests to demonstrate the reliability of the numerical simulations for frequencies, which, as explained in \secref{results}, lie between 1.5 and 2.5~MHz. First, we did a mesh convergence analysis choosing a pyrex lid of height $h_\mathrm{gl}=1100~\SImum$ and a centered piezo transducer ($d_\mathrm{pz}=0~\SImum$). We use meshes of triangular elements with a coarseness controlled by the mesh resolution parameter $m_\mathrm{res}$ between zero and unity. As illustrated in \figref{mesh} for $m_\mathrm{res}=0.35$, the maximum linear size of the mesh elements is set to be $m_\mathrm{res}\:h_\mr{ch}$ in the channel, $0.1\: m_\mathrm{res}\:h_\mr{ch}$ at six corners and $1.4\: m_\mathrm{res}\:h_\mr{ch}$ elsewhere. We considered the four mesh resolutions $m_\mathrm{res}=0.10$, $0.14$, $0.20$ and $0.35$, corresponding to a doubling in the number of mesh elements for each step, and then used the results for the finest mesh $m_\mathrm{res}^* = 0.10$ as the reference solution. For a given field $F(f,m_\mathrm{res})$, we define the mesh- and frequency-dependent relative error $\mr{err}\big[F(f,m_\mr{res})\big]$ and its average $\overline{\mathrm{err}}\big[F(m_\mathrm{res})\big]$ over $N$ discrete frequencies $f_k$ by
 \bsubal
 \eqlab{relerr}
 \mathrm{err}\big[F(f,m_\mathrm{res})\big]
 & =
 \left|\frac{F(f,m_\mathrm{res})-F(f,m_\mathrm{res}^*)}
 {F(f,m_\mathrm{res}^*)}\right|,\\[2mm]
 \eqlab{error}
 \overline{\mathrm{err}}\big[F(m_\mathrm{res})\big]
 &=\frac{1}{N}\sum_{n=k}^{N}
 \mathrm{err}\big[F(f_k,m_\mathrm{res})\Big].
 \esubal

In \figref{errorfig}(a) are shown plots of the average relative errors of the Lagrangian $L$  and the Hamiltonian $H$ versus the mesh resolution $m_\mathrm{res}$ for $N = 1001$ frequencies between 1.5 and 2.5~MHz in steps of 1~kHz. This error is below 1~\% for all meshes and decreases proportional to $m_\mathrm{res}^2$ as $m_\mathrm{res} \rightarrow 0$. In \figref{errorfig}(b) is plotted the relative error for each of the $N$ frequencies for a fixed mesh resolution $m_\mr{mes}=0.14$. The relative error fluctuates as a function of frequency with pronounced local maxima near what turns out to be resonance frequencies. The largest maximum is $5\times 10^{-3}$ at $f\simeq 1.7$~MHz, which is an order of magnitude larger than the average relative error of $3\times 10^{-4}$, but still below 1~\%. To trade off between computational time and accuracy, we use the mesh resolution $m_\mr{res} = 0.14$ (yielding 150,000 degrees of freedom) in all the numerical simulations here presented.

\begin{figure}[t]
\begin{picture}(233,144)
\put(-55,-20){
\includegraphics[width=10cm]{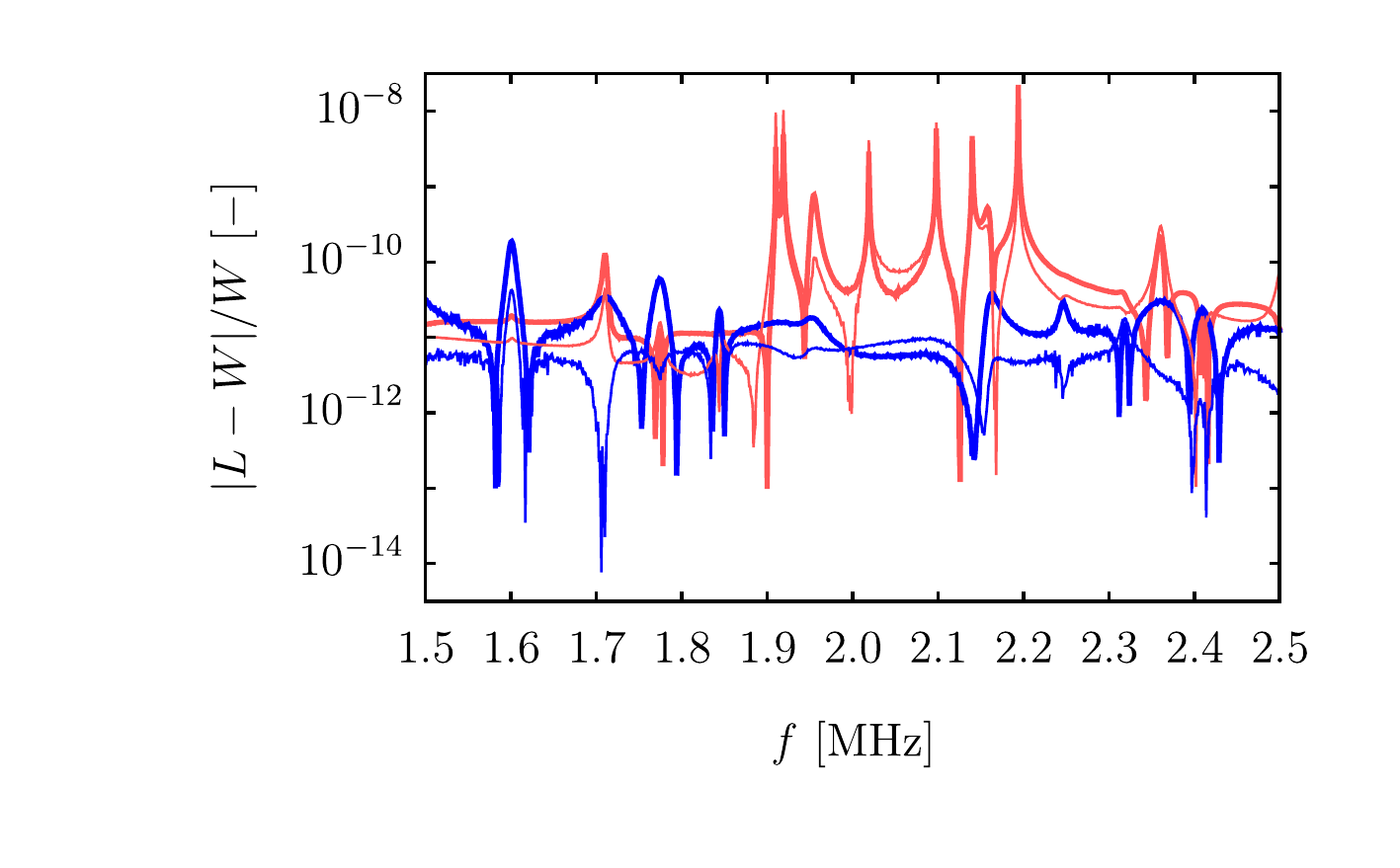}}
\end{picture}
\caption{\figlab{modelcheckfig}
Relative deviations for the real part (light red lines) and imaginary part
(blue lines) of the Lagrangian $\hat{L}$ from the external work
$\hat{W}$ versus frequency $f$ for
$d_\mathrm{pz}=0~\SImum$,
$h_\mathrm{gl}=1100~\SImum$ for two mesh resolutions:
$m_\mathrm{res}=0.10$ (thin lines) and $m_\mathrm{res}=0.14$
(thick lines).}
\end{figure}

\begin{figure*}[]
\center{\includegraphics[]{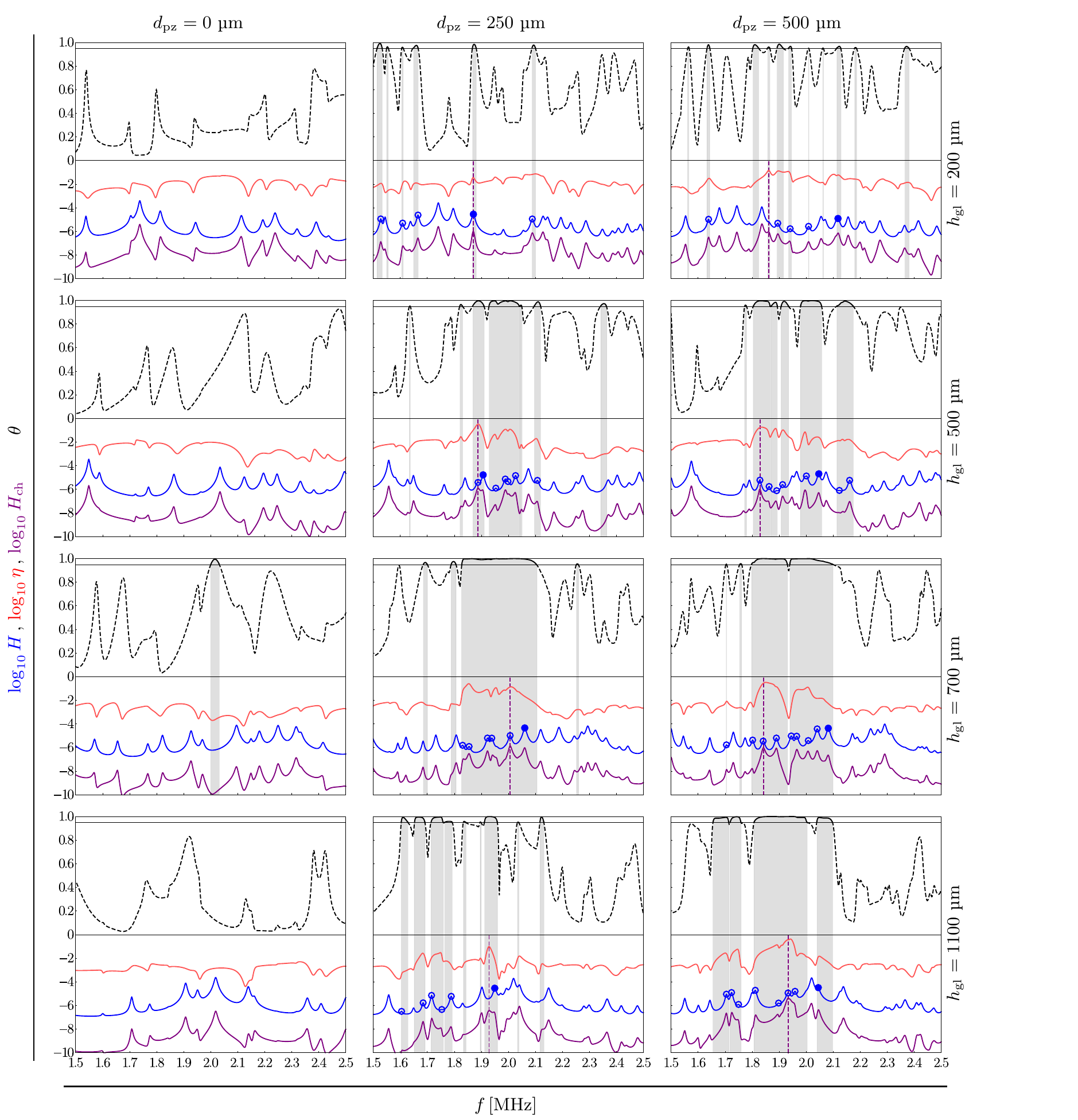}}
\caption{\figlab{HYPpyrex}
Plots of the indicators versus the frequency $f$ between 1.5~MHz and 2.5~MHz in a pyrex-silicon device: the acoustophoretic orientation $\theta$ (black/black dashed lines), the Hamiltonian $H$ (dark blue lines), the acoustofluidic yield $\eta$ (light red lines) and the channel energy $H_\mathrm{ch}$ (purple line), each shown for the three positions $d_\mathrm{pz}$ (values on the top of columns) of the piezo transducer and four glass heights (values on the right of rows). Gray-shaded regions are the ACP-bands \eqref{apbandeq} for $\theta=0.95$. On the graphs we have highlighted the local maxima in $H(f_n)$ ($\circ$), the global maximum $H(f_H)$ ($\bullet$) and the optimal acoustophoretic frequency $\facp$ (vertical dashed line).}
\end{figure*}

The internal consistency in the implementation of the governing equations and the boundary conditions has been tested by using \eqref{lwequiv}, which states that the total work $\hat{W}$ provided by the applied potential $\phi_\mr{app}$ on the piezo transducer must equal the total Lagrangian $\hat{L}$ defined in Eq.~\eqnoref{Ltotal}. In \figref{modelcheckfig} the relative deviation $|(\hat{L}-\hat{W})/\hat{W}|$ between these two quantities is plotted as function of the actuation frequency for the two mesh resolutions $m_\mathrm{res}=0.10$ and $0.14$, and it is found to be numerically zero for all of the frequencies.

\section{Results}
\seclab{results}

Since the usual standing half-wave resonance in the microchannel is $f_0=\frac{c}{2 w_\mathrm{ch}} = 1.974$~MHz, we simulate the response of the acoustofluidic device to actuation frequencies from $1.5$ to $2.5$~MHz. We vary the lid height $h_\mr{gl}$ and the displacement $d_\mr{pz}$ of the piezo transducer as indicated in \tabref{geometry}, and we study two different transparent materials for the glass lid, namely pyrex and the more stiff and heavy aluminium oxynitride (ALON) (see Supplementary Material at  for details).

\subsection{Procedure for identifying good acoustophoresis}
\seclab{idproc}
In experiments aiming to the separation or focusing of particles, a strong and well-oriented acoustophoretic force is necessary to provide the adequate displacement of the particle stream towards the pressure node fast enough to be completed before the particle stream leaves the device. The occurrence of a resonant state is therefore a necessary condition to produce a strong acoustophoretic force. In order to identify the optimal actuation frequency $f$ for a given device geometry and applied voltage $\phi_\mathrm{app}$, we employ the indicators introduced in \secref{encharsec}.

First, we locate the frequencies for which the acoustophoretic mean orientation $\theta$, as defined in  Eq.~\eqnoref{mopw}, exceeds a threshold value \mbox{$0<\theta_0<1$} close to one. Therefore, we introduce the concept of \emph{acoustophoretic bands} (ACP-bands) as
 \bal
 \eqlab{apbandeq}
 &\text{ACP-bands = frequencies for which $\theta(f) > \theta_0$}.
 \eal
This requirement ensures that the orientation of the pressure field and the resulting acoustic radiation force leads to microparticle migration towards a vertical nodal line, and can be regarded as a quality measure of the acoustic force.

On the other hand, the acoustophoretic force intensity is regulated by the amount of energy present in the channel for a specific actuation frequency. Therefore, we look for frequencies for which either the system energy $H$ or the channel energy $H_\mathrm{ch}=\eta\,H$ attain maxima. In the first case, the actuation frequencies correspond to resonance frequencies of the entire system, while in the second case we only demand for a sufficient amount of energy in the channel to obtain acoustophoresis. Note that channel resonances and system resonances do not need to coincide. Note also that the detection of a channel resonance is achievable only by indirect measurements, such as $\mu$-PIV (micro-particle image velocimetry) experiments, while the detection of a system resonance is easily achievable by means of electrical measurement, for example by looking at the minima of the real part of the impedance.

We can now establish a numerical procedure for identifying effective actuation frequencies in terms of acoustophoresis as follows: (i)~calculate the indicators  $\theta$, $\eta$, $H_\mathrm{ch}$ and $H$ as a function of frequency $f$ in a chosen interval, (ii)~identify the ACP-bands for a threshold value, (iii)~locate within the ACP-bands the frequencies $f_n$ leading to local maxima of $H$ and (iv) compute the optimal actuation frequency $\facp$ defined as
 \beq{facpDef}
 \facp=f\,\,\text{that maximizes } H_\mathrm{ch}\,\,\text{in the ACP-bands.}
 \eeq
For comparison purposes, we also keep track of the frequency $f_H$ for which $H(f_n)$ has a global maximum within the acoustophoretic bands.

\subsection{Mechanical characterization}
\seclab{PyrexSiliconDevices}
An example of the identification procedure for pyrex/silicon devices is reported in \figref{HYPpyrex}. Here, $\theta$, $\eta$, $H_\mathrm{ch}$ and $H$ are plotted versus the actuation frequency $f$ in the interval from 1.5 to 2.5~MHz for all the twelve combinations of glass height and piezo displacement as listed in \tabref{geometry}.
For each device, the top graph contains the lin-lin plot of $\theta$ (black/black-dashed line), while the bottom graph contains the log-lin plots of the acoustofluidic yield $\eta$ (red line), the channel energy $H_\mathrm{ch}/(1~\SIPa)$ (purple line) and the total energy $H/(1~\SIPa)$ (blue line). We show the ACP-bands for $\theta_0 = 0.95$ (gray areas), the local maxima of the total energy within the ACP-bands $H(f_n)$ ($\circ$), the global maximum of the total energy within the ACP-bands $H(f_H)$ ($\bullet$), and finally the optimal frequency $\facp$ (vertical dashed line).

Some general features can be observed in the figure. Firstly, the ACP-bands are practically absent for the symmetrically placed transducer ($d_\mr{pz} = 0$), and the orientation $\theta$ rarely surpasses even $0.9$; this is because the specific transducer, that is Pz26, is mainly characterized by a compression/extension actuation mode in the $z$-direction \cite{Ferroperm_2015}, that in a piezo/chip symmetric configuration does not provide a sufficient excitation for the $y$-directed vibrational modes that are those responsible for the acoustophoresis.
Secondly, for the asymmetrically placed transducer, the width of the ACP-bands increases for increasing lid height $h_\mr{gl}$. Thirdly, we note that not for all of the cases we investigated, the optimal acoustophoretic frequency $\facp$
coincides with a local maximum in $H$ (see the cases $d_\mathrm{pz}=250~\SImum$, $h_\mathrm{gl}=1100~\SImum$ and $d_\mathrm{pz}=500~\SImum$, $h_\mathrm{gl}=200~\SImum$). These observations point out to the complex interplay between the different parts of the device and highlight the fact that the channel, where acoustophoresis is observed, is a small part of the system, and it can contain a remarkable amount of energy even when the whole system is not resonating.

\begin{figure}[!!t]
\includegraphics[width=\columnwidth]{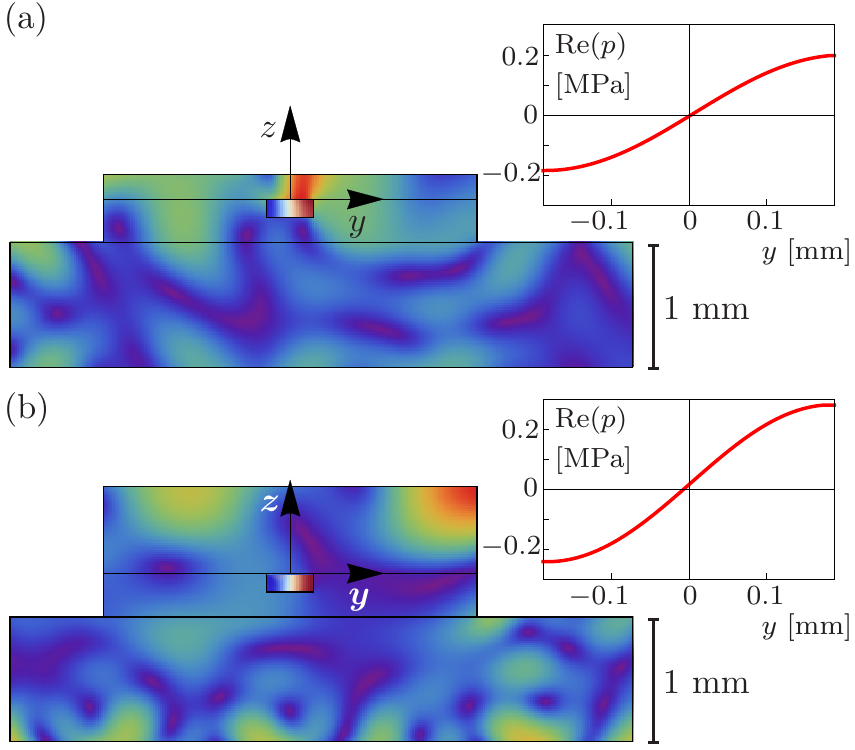}
\caption{\figlab{coloredfig1}
Color plot of the magnitude $|\re(\bm{u})|$ of the real part of the displacement field in the
solids (rainbow colormap) and the real part $\re(p)$ of the acoustic pressure  in the fluid (red-white-blue colormap) for two silicon-pyrex devices with (a)
$h_\mathrm{gl}=200~\SImum$ at $f = \facp = 1.875\,\mathrm{Mhz}$, and (b)
$h_\mathrm{gl}=700~\SImum$ at $f = \facp = 2.099\,\mathrm{Mhz}$. The insets show the real part $\re(p)$ of the pressure  along the horizontal centerline of the water-filled microchannel.
}
\end{figure}

\begin{figure*}[!!t]
\hspace*{-2cm}\centerline{\includegraphics[]{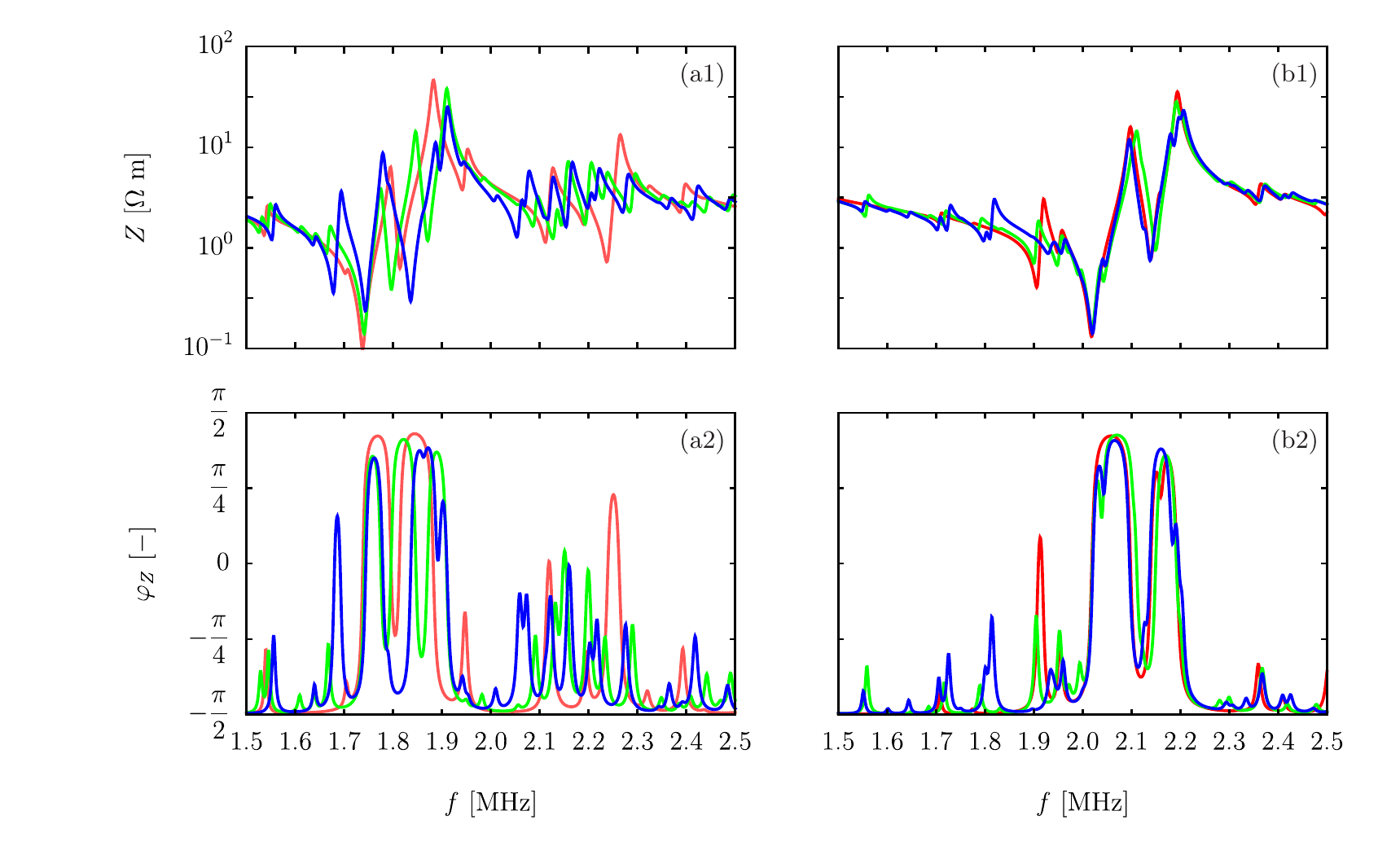}}
\caption{\figlab{ZPfigure1}
Magnitude $Z$ and phase $\varphi_Z$ of the electric impedance as function of the frequency $f$ for pyrex-silicon devices with three different off-center displacements of the piezo transducer: $d_\mathrm{pz}=0~\SImum$ (red), $250~\SImum$ (green), and $500~\SImum$ (blue).
(a1)--(a2) $Z$ and $\varphi_Z$, respectively, for devices with lid thickness $h_\mathrm{gl}=200~\SImum$. (b1)--(b2) $Z$ and $\varphi_Z$, respectively, for devices with lid thickness $h_\mathrm{gl}=1100~\SImum$.}
\end{figure*}

\subsection{Example of identifying good acoustophoresis}
The ability to identify optimal frequencies for acoustophoresis by using the procedure above introduced, is illustrated by the two examples shown in \figref{coloredfig1}. The two systems have pyrex lids and nearly identical geometries defined by the parameters listed in \tabref{geometry}. The piezo displacement is $d_\mr{pz} = 250~\SImum$ in both cases, while the glass lid height is $h_\mathrm{gl}=200~\SImum$ (\figref{coloredfig1}a) and $h_\mathrm{gl}=700~\SImum$ (\figref{coloredfig1}b). The respective optimal actuation frequencies $\facp = 1.875$ and $\facp=2.099$~MHz have been obtained by using the definition \eqref{facpDef}.
Despite the very different displacement fields in the surrounding chip material, which is represented as the magnitude of the real part of the displacement $|\re(\bm{u})|$ by the color plot (from blue zero, via green, to red maximum), both systems exhibit a nearly perfect horizontal standing pressure wave $\re(p)$ (color plot from blue minimum, through white, to red maximum or line plot in the inset) in the fluid. This occurrence has been observed experimentally \cite{Barnkob_2010, Augustsson_2011, Barnkob_2012}, and it is ideal for forcing the particles to the vertical pressure nodal plane in the center of the channel. By definition of $\facp$, the orientation $\theta$ is close to unity in both systems. In all of the cases for which we have located $\facp$ by using the identification procedure, we have observed similar behaviors in the pressure field inside the microchannel.

\subsection{Electrical characterization}
So far, the device characterization has only involved the mechanical indicators $\theta$, $H$, $H_\mathrm{ch}$ and $\eta$, that unfortunately cannot be measured directly and easily during the experiments. It would be advantageous if the acoustophoretic performance of a given system could be assessed by measuring only electrical quantities through the connection of the piezo transducer. A good candidate for an easily accessible indicator is the complex-valued electrical impedance \mbox{$\hat{Z}=Z\,e^{i\varphi_Z}$} in \eqref{impdef}, which can be measured experimentally by the use of an impedance analyzer. This frequency-dependent complex-valued indicator can be written as $\hat{Z}= R + \ii X$, where the real part $R$ is the resistance, and the imaginary part $X$ is the reactance. A system is considered to behave inductively for $X > 0$ (or $0 < \varphi_Z < \tfrac{\pi}{2}$) and capacitively for $X < 0$ (or $-\tfrac{\pi}{2} < \varphi_Z < 0$). Depending on the internal structure of the electric circuit (RLC series or parallel), driving the system at the resonance frequency can amplify voltage or current \cite{Nilsson_2014}.
This occurs for zero reactance when the system is purely resistive, and the current is in phase with the applied voltage. In this situation, the magnitude of the impedance can either be at a local minimum or at a local maximum, and the system is said to be in a resonant or anti-resonant state, respectively. Driving the device at a resonance frequency, leads to a local maximum of stored energy $H$, and this value can be large even for a low input power.

In \figref{ZPfigure1} are shown the frequency dependence of the magnitude $Z$ (top row, unit $\Omega\:\mathrm{m}$ as this is a 2D calculation) and the phase $\varphi_Z$ (bottom row) of the electrical impedance $\hat{Z}$ for silicon-pyrex devices with lid-height $h_\mr{gl} = 200~\SImum$ (first column), $1100~\SImum$ (second column) and three different displacements $d_\mr{pz} = 0~\SImum$ (red), $250~\SImum$ (green), and $500~\SImum$ (blue) of the piezo transducer. The response of the system alternates between resonances and anti-resonances (minima and maxima in the impedance while the phase crosses zero), but for the chosen parameters, the system has a prevalent capacitive behavior (\mbox{$\varphi_Z<0$}) with intermittent inductive behavior (\mbox{$\varphi_Z>0$}). We observe that in the case \figref{ZPfigure1}(a1) and (a2) of the smaller chip, where the piezo transducer is weakly loaded, the impedance exhibits many small fluctuations as a function of frequency and is sensitive to the off-center position of the piezo transducer. For the larger chip \figref{ZPfigure1}(b1) and (b2), less fluctuation and less sensitivity to $d_\mr{pz}$ is observed: the impedance and the phase curves is nearly independent of $d_\mr{pz}$ and the resonance frequencies nearly coincide as does those of the anti-resonances.
These plots confirm the picture regarding the complicated interactions between the different parts of the system and the great sensitivity of the resonance conditions to the possible geometric configurations for the system.

\subsection{Procedure for identifying resonances from electrical impedance}

Alongside these qualitative comments, it is important to establish to what extent a criterion based on the electrical impedance characteristics can identify good acoustophoresis frequencies.
\textit{A priori} it is not possible to provide a criterion to detect frequencies (resonance or anti-resonance) leading to a high energy content for the system.
Instead, we try to establish empirical rules by direct comparison of the mechanical indicators and the electrical impedance.
\begin{figure}[!!t]
\centerline{
\hspace*{-0.8cm}\includegraphics[width=\columnwidth]{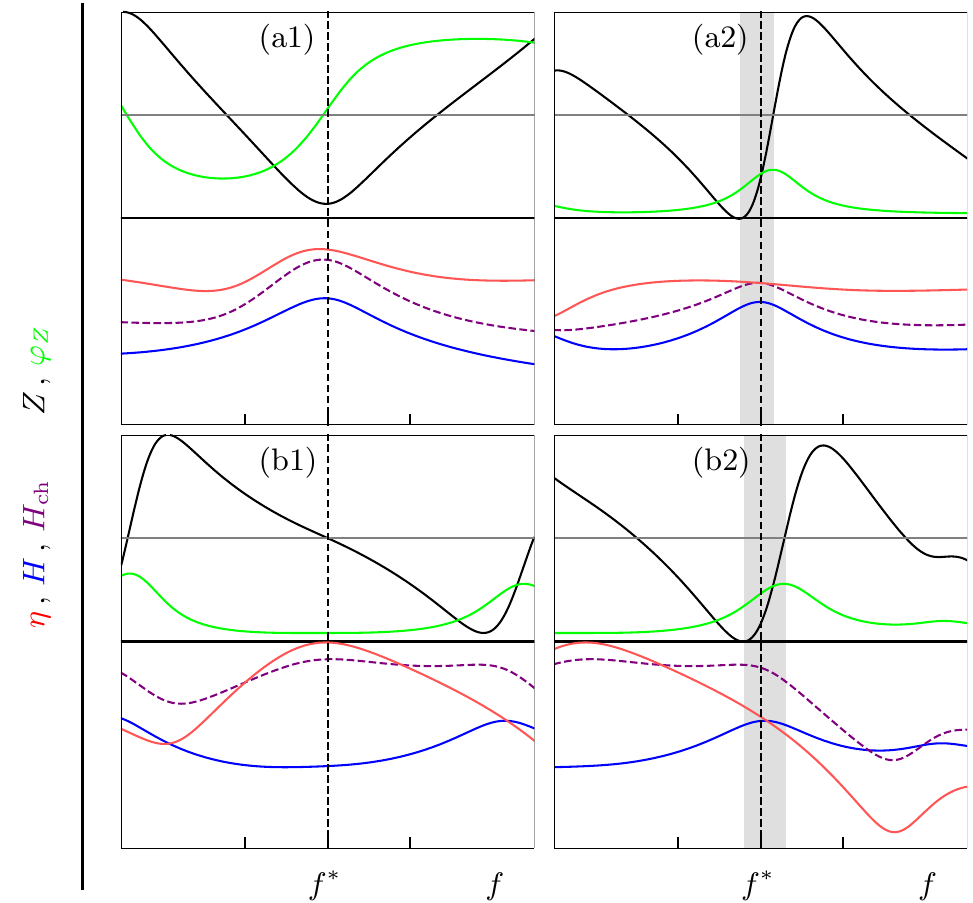}}
\noindent
\caption{\figlab{elecVSmech}
The magnitude $Z$ (black, arbitrary scale) and the phase $\varphi_Z$ (green, thin horizontal line \mbox{$\varphi_Z=0$}) of the impedance $\hat{Z}$, the energy $H$ (blue, arbitrary scale), the yield $\eta$ (red, arbitrary scale), and the channel energy $H_\mathrm{ch}$ (purple, arbitrary scale) as a function of frequency $f$ in an interval characterized by $\Delta f = 10~\SIkHz$ around a frequency $f^*$ at which the channel energy is a maximum.
(a1) Chip with $h_\mathrm{gl}=200~\SImum$, $d_\mathrm{pz}=250~\SImum$ and $f^*=f_n=\facp=1.870$~MHz.
(a2) Same as (a1) but at $f^*=f_n=1.666$~MHz.
(b1) Chip with $h_\mathrm{gl}=1100~\SImum$, $d_\mathrm{pz}=250~\SImum$ and $f^*=\facp=1.929$~MHz.
(b2) Same as (b1) but at $f^*=f_n=1.950$~MHz. The gray area is the interval between the frequencies given by \eqref{impcrit2} (case (b1) cannot be detected).}
\end{figure}

In \figref{elecVSmech} four prototypical cases observed in the behavior of the mechanical and the electrical indicators are illustrated. The weakly loaded piezo transducer (smallest height of the lid) is shown in \figref{elecVSmech}(a1). Here a situation corresponding to a classical resonance condition for a RLC series circuit (\mbox{$\varphi_Z=0$} and \mbox{$Z=\mathrm{min}$}) is illustrated, and the optimal acoustophoretic frequency coincides with the frequency at which $H$ attains a maximum, that is $\facp=f_H$. This case is different from the behavior observed for the same system but for a generic resonance $f_n\neq \facp$ (a2), where the resonance frequency is in between a maximum in the phase and a minimum in the impedance magnitude. The extremum in the phase is a maximum because the system mostly behaves as a capacitor (\mbox{$\varphi_Z<0$}) and combining the magnitude of the impedance and the phase it has the effect to minimize the resistive part of the impedance thus minimizing the power input.
This phenomenology can be observed also for the other generic resonance $f_n\neq \facp$ (even outside the ACP-bands) also in the case when the piezo is heavily loaded (b2), namely for $h_\mathrm{gl}=1100~\SImum$.
Analogous behavior has been observed but less frequently when the phase attains a positive maximum.

Finally, a quite peculiar behavior has been observed for the heavily loaded piezo (b1) when $f=\facp$ but $f\neq f_n$. In this case there is no evidence for a resonance in the system, no zero/maximum in the phase neither a minimum in the impedance, while the energy content in the channel has a maximum.
The pairing of this observation with the analysis of the energy quantification in \figref{elecVSmech} reveals that in this situation the channel energy is slightly higher than that achievable for the resonance frequency $f_n=1.950$~MHz, because the yield is quite high. This peculiar case can be seen as a case in which the channel is resonating while the whole system is not, and it is the most difficult to detect electrically since there is no trace of the energy content of the channel in the impedance measurements.

Based on these observations, a criterion to detect resonance frequencies $f_{Z,n}$ for the system can be constructed as follows
\begin{align}\label{eq:impcrit2}
f_{Z,n}&\;\text{is in the interval between}\;f_{Z}\;\text{and}\;f_{\varphi_Z}\;\text{where}\\
\text{(i)}\;&f_{Z}=\tilde{f}\,,\tilde{f}\;\text{minimizes}\;Z\;\text{and}\,,\nonumber\\
\text{(ii)}\;&f_{\varphi_Z}=\tilde{f}\;\left\{\begin{array}{l}
\text{either}\;\varphi_Z(\tilde{f})=0\;\text{and}\;\varphi_Z'(\tilde{f})>0\\
\text{or minimizes}\;|\varphi_Z(\tilde{f})|\,.
\end{array}\right.\,\nonumber
\end{align}
Note that this criterion is a less restrictive version of the criterion used to detect resonance frequencies for a RLC series circuit, equipped with the additional condition that the phase attains a negative maximum or a positive minimum. The criterion \eqref{impcrit2} does not provide the location of an exact resonance frequency of the system, but it addresses a frequency interval in which is possible to find the resonance frequency. The intervals are formally given by \mbox{$f_{Z}<f_{Z,n}<f_{\varphi_Z}$} and \mbox{$f_{\varphi_Z}<f_{Z,n}<f_{Z}$} for \mbox{$\varphi_Z<0$} and \mbox{$\varphi_Z>0$}, respectively. As shown in the figure, these frequency intervals are quite narrow and for the classical resonance conditions \mbox{$\varphi_Z=0$} the frequency bounds coincide and a single frequency is recovered.

\begin{figure}[!!b]
\vspace*{-0.3cm}
\hspace*{-1.8cm}\includegraphics[angle=90]{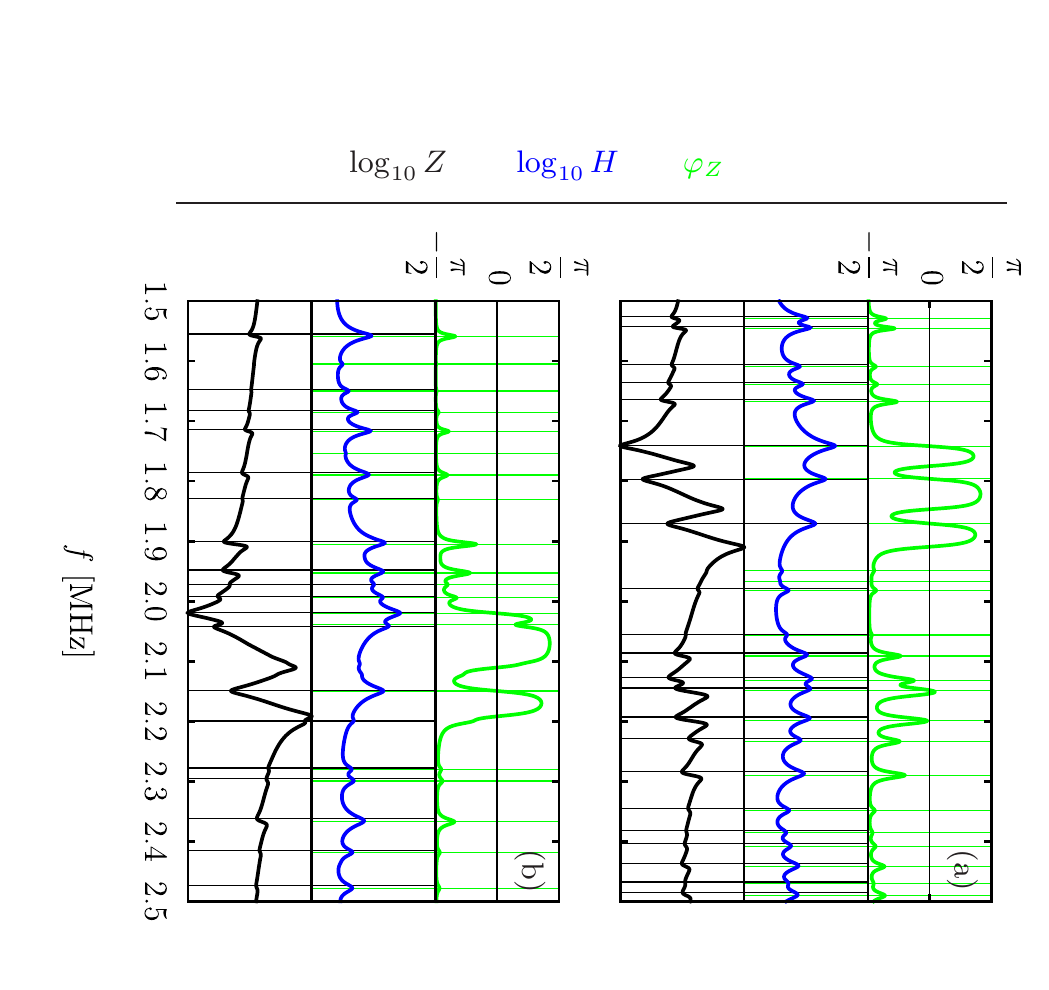}
\vspace*{-0.3cm}
\caption{\figlab{ea}Phase $\varphi_Z$ (green), impedance $Z$ (black) and energy $H$ (blue) as function of the frequency $f$ for pyrex-silicon devices with $d_\mathrm{pz}=250~\SImum$ for (a) $h_\mathrm{gl}=200~\SImum$ and (b) $h_\mathrm{gl}=1100~\SImum$. Vertical black and green lines mark the frequencies obtained via the criteria \eqref{impcrit2} (i) and (ii), respectively.}
\end{figure}

\figref{ea} reports the results of the application of the criterion \eqref{impcrit2} to the detection of resonance frequency intervals for two silicon-glass device with \mbox{$d_\mathrm{pz}=250~\SImum$} for (a) $h_\mathrm{gl}=200~\SImum$ and (b) $h_\mathrm{gl}=1100~\SImum$, by depicting the spectral behavior of the phase $\varphi_Z$ (top plot), the energy $H$ (center plot) and the impedance $|\hat{Z}|$ (bottom plot). The green and black vertical lines mark the frequency bounds obtained through \eqref{impcrit2} by applying the criteria (i) and (ii), respectively.
Here the effectiveness of the criterion \eqref{impcrit2} in the detection of frequency intervals containing resonance frequencies can be appreciated. Indeed, in almost all of the cases the energy maxima have been enclosed in the predicted intervals. It can also be observed that when one of the two criteria is missing (mostly because of the procedure used in the extremum calculation over numerical data), the other can still provide a good approximation for the resonance frequencies.
Despite the criterion \eqref{impcrit2} can detect resonance frequencies for the system, the effectiveness of these frequencies in producing particle focusing is still hidden by the fact that the impedance indicates an overall performance of the system and does not contain information about the orientation of the pressure wave in the microchannel.

\section{Conclusion}
\seclab{conclusion}
The Helmholtz equations governing the elastic, acoustic and piezoelectric waves in an acoustophoretic device have been formulated in terms of the free Lagrangian densities of each sub-system: the piezo-transducer, the silicon substrate, the glass lid and water-filled microchannel. The sub-systems have been coupled and boundary conditions have been imposed by adding surface contributions to the free Lagrangian densities. In this way, we have constructed a consistent model taking into account for the coupling between the different sub-systems and the electric actuation. The model have been implemented and thoroughly tested by means of numerical simulations by using the weak-form implementation in the commercial software Comsol Multiphysics v4.4a.

Three frequency-dependent mechanical indicators were introduced to characterize the acoustophoretic response of the system: the total energy $H$, the acoustofluidic yield $\eta$, and the mean acoustophoretic orientation $\theta$. A specific procedure that allows for the identification of optimal frequencies $\facp$ has been provided, so that it is possible to obtain good acoustophoretic focusing of suspended microparticles flowing through the microchannel. Additional electrical indicators, such as the impedance $\hat{Z}$ of the system, the power input $P$ and the Q-value were correlated with the Lagrangian and the Hamiltonian of the system.

We have (i)~illustrated the general guideline to obtain optimal actuation frequencies $\facp$ from spectral analysis of the mechanical indicators and (ii)~by comparing mechanical and electric indicators, we provided an empirical criterion to detect resonance frequencies (high energy in the system) by using impedance characteristics.

To exemplify the use of the present model as a possible design tool, we have analyzed the variations in the spectral behavior induced by changing the commonly used glass lid from pyrex to ALON (see Supplementary Material at for details).
The analysis has revealed unique features of ALON such as stabilization of the optimal resonance frequency as the height of the lid increases. This behavior deserves for further experimental studies employing other stiff materials such as by sapphire \cite{Oliver_1992}, for which some perspective in the bonding process are known \cite{Manasevit_1964}. The rule \eqref{impcrit2} based on the comparative analysis of the mechanical and electrical characteristics of the device is easy to implement in the applications and it can successfully predict the frequency ranges in which system resonances occur

In summary, the model we reported may be employed as (i)~an optimal design tool in the future development of acoustophoretic devices or (ii)~to narrow experimental frequency intervals for the detection of optimal/resonance frequencies by using optical techniques.
It demonstrates the effectiveness of the employment of basic theory in the area of coupled mechanical and electrical systems, and as such is closely related to the developments in mechatronics \cite{Preumont_2006, Lenk_2012}, and it points towards the possibility of reducing the expensive and time-consuming trial-and-error fabrication procedures presently being used in the field.

\subsection*{Acknowledgement}
This work was supported by the Knut and Alice Wallenberg Foundation (Grant No. KAW 2012.0023).

\begin{table}[!!t]
\caption{\tablab{symbols}
Summary of symbols and their meanings.
Scalars are addressed by means of plain roman and Greek letters.
Script symbols refers to Lagrangian, Hamiltonian, work and power densities.
Vectors and tensors of 2nd, 3rd and 4th order are addressed by using bold roman,
bold Greek, bold capital roman and bold capital Greek, respectively.
Prime-ed symbols address the Voigt notation for the simmetrizable tensors.
}
\begin{tabular}{ccp{4.5cm}p{1.5cm}}
\toprule
Order & Symbol & Description & Units\\
\cline{1-4}
0            & $f$ & Frequency & $\mathrm{Hz}$ \\
             & $\omega$ & Angular frequency & $\mathrm{rad/s}$ \\
             & $\rho$ & Density & $\mathrm{kg/m^3}$ \\
             & $E$ & Young Modulus & $\mathrm{N/m^2}$ \\
             & $\nu$ & Poisson Ratio & $1$ \\
             & $p$ & Pressure & $\mathrm{Pa}$ \\
             & $\phi$ & Electric potential & $V$ \\
             & $\mathscr{L}$ & Lagrangian density & $\mathrm{J/m^3}$ \\
             & $\mathscr{H}$ & Hamiltonian density & $\mathrm{J/m^3}$ \\
             & $\mathscr{W}$ & Work density & $\mathrm{J/m^3}$ \\
             & $\mathscr{P}$ & Power density & $\mathrm{J/s\cdot m^3}$ \\
             & $L$ & Lagrangian & $\mathrm{J}$ \\
             & $H$ & Hamiltonian & $\mathrm{J}$ \\
             & $W$ & Work & $\mathrm{J}$\\
             & $P$ & Electric power &  $\mathrm{J/s}$\\
             & $Z$ & Electric impedance & $\mathrm{\Omega}$\\
             & $I$ & Electric current & $\mathrm{A}$\\
             & $\eta$ & Acoustofluidic yield & $1$\\
             & $\theta$ & Acoustophoretic orientation & $1$\\
1            & $\bm{u}$ & Displacement & $\mathrm{m}$ \\
             & $\bm{v}$ & Velocity & $\mathrm{m/s}$ \\
             & $\bm{a}$ & Acceleration & $\mathrm{m/s^2}$ \\
             & $\bm{d}$ & Electric displacement & $\mathrm{C/m^2}$ \\
             & $\bm{\hat{n}}$ & Outward normal & $1$ \\
2            & $\bm{1}$ & Unit tensor & $1$ \\
             & $\bm{\varepsilon}$ & Dielectric tensor & $\mathrm{F/m^2}$\\
             & $\bm{\gamma}$ & Strain $\frac{1}{2}(\nabla\bm{u}+\bm{u}\nabla)$
             & $-$ \\
             & $\bm{\sigma}$ & Stress & $\mathrm{N/m^2}$ \\
3            & $\bm{P}$ & Electric Field-Strain & $\mathrm{C/m^2}$ \\
             &          & piezoelectric coupling & \\
4            & $\bm{\Sigma}$ & Stiffness tensor & $\mathrm{N/m^2}$ \\
\botrule
\end{tabular}
\end{table}

\appendix

\section{Axial modes analysis}
\seclab{2D3Danalysis}
For completeness, we  use an analysis of the axial modes to briefly discuss what is the connection between the two dimensional (2D) model presented in the manuscript and a full three dimensional (3D) model. \tabref{symbols} contains the symbols used here as well as in the main manuscript.

Using the combined variable $\qqq = [\uuu,p,\phi]^{\mr{T}}$, the governing equations~\eqnoref{neeq}, \eqnoref{tdaweq} and \eqnoref{tdpeq} can be obtained from the Euler--Lagrange equation with the Lagrangian density
 \beq{Lagrange3D}
 \mathscr{L}(\qqq,\nablabf \qqq)=
 \qqq^\dagger\mathbf{K}\qqq
 +(\nablabf \qqq)^\dagger\mathbf{M}\nablabf \qqq,
 \eeq
where $\qqq^\dagger=(\qqq^*)^{\mr{T}}$, $\qqq^\dagger\qqq$
is the inner product for the space of the complex valued states,
$\mathbf{K}$ is a matrix with scalar entries
 \begin{equation}
 \mathbf{K}=
 \left(
 \begin{array}{ccc}
 \rho\omega^2 & 0 & 0 \\
 0 & -\kappa_s & 0 \\
 0 & 0 & 0
 \end{array}
 \right),
 \end{equation}
and $\mathbf{M}$ is a matrix with tensorial entries
\begin{equation}
\mathbf{M}=
\left(
\begin{array}{ccc}
\bm{\Sigma} & 0 & -\bm{P} \\
0 & (\rho\omega^2)^{-1} & 0 \\
-\bm{P}^{\mr{T}}& 0 & \bm{\varepsilon}
\end{array}
\right).
\end{equation}

Let the position coordinate be written as $\rrr = (x, \rrr_\perp)$, such that $x$ is the axial coordinate and $\rrr_\perp = (y,z)$ is the cross section coordinates. We assume that the device is axially periodic along the $x$ axis with period $\ell$ so that $0 < x < \ell$ and $\qqq(0,\rrr_\perp,\omega) = \qqq(\ell,\rrr_\perp,\omega)$. Using a discrete Fourier series in the axial coordinate, we can write
 \begin{equation}
 \qqq(x,\rrr_\perp,\omega)= \sum_{n=-\infty}^{\infty}
 \hat{\qqq}_n(\rrr_\perp,\omega)\: \ee^{\ii 2\pi nx/\ell},
 \end{equation}
where the amplitudes $\hat{\qqq}_n$ are given by
 \begin{equation}
 \hat{\qqq}_n(\bm{x}_\perp,\omega)=
 \frac{1}{\ell}\int_{-\ell/2}^{\ell/2}\qqq(\bm{x}_\perp,\omega)\:\ee^{-\ii 2\pi nx/\ell}
 \:\dm x.
 \end{equation}
The nabla operator $\hat{\nablabf}$ acting on $\hat{\qqq}$ is written as
 \begin{equation}
 \hat{\nablabf} = \ii\:\frac{2\pi n}{\ell}\een_x + \nablabf_\perp,
 \end{equation}
where $\nablabf_\perp = \een_y\:\pp_y + \een_z\:\pp_z$ is the cross-sectional nabla operator. Since the quadratic Lagrangian density~\eqnoref{Lagrange3D} leads to linear equations of motion, the axial-mode decomposition of any solution $\qqq$ is unique and complete, and each of the axial modes $\hat{\qqq}_n$ can be determined from their respective Lagrangian density
 \begin{equation}
 \mathscr{L}_n(\hat{\qqq}_n,\bm{\hat{\nabla}}\hat{\qqq}_n)=
 \hat{\qqq}_n^\dagger\mathbf{K}\hat{\qqq}_n+
 (\bm{\hat{\nabla}}\hat{\qqq}_n)^\dagger
 \mathbf{M}\bm{\hat{\nabla}}\hat{\qqq}_n.
 \end{equation}
This splitting is possible because $\mathbf{M}$ and $\mathbf{K}$ does not depend on the spatial coordinates.

For the axially invariant mode $n=0$, there is no dependence on $x$ and $\hat{\nablabf} = \nablabf_\perp$. Moreover the displacement field is restricted to be in the cross section, $\uuu\cdot\een_x = 0$. Consequently, the Lagrangian density $\mathscr{L}_0$ for the $n=0$ mode can be written as
 \begin{equation}
 \mathscr{L}_0(\hat{\qqq}_0,\nablabf_\perp\hat{\qqq}_0)=
 \hat{\qqq}_0^\dagger\mathbf{K}\hat{\qqq}_0+
 (\nablabf_\perp\hat{\qqq}_0)^\dagger \mathbf{M}_\perp
 \nablabf_\perp\hat{\qqq}_0,
 \end{equation}
where $\mathbf{M}_\perp$ only contains the tensorial components in the cross section plane from $\mathbf{M}$. This Lagrangian density directly leads to the 2D equations of motion with bulk material properties studied in \secref{model}. In an extended treatment, the axial modes can be used to approach a full 3D solution by taking higher modes with $|n| > 0$ into account.

\bibliography{acoustofluidics2}

\end{document}